\documentclass{article}

\usepackage{graphicx} 
\usepackage{amsmath,amsfonts,amssymb,amsthm}
\usepackage[a4paper]{geometry}
\usepackage{natbib}

\usepackage{booktabs}
\usepackage[pdftex,pdfview={Fit}, pdfstartview={Fit}]{hyperref}
\usepackage{multirow}
\usepackage{xcolor}
\usepackage{longtable}
\usepackage{threeparttable}
\usepackage{array}
\newcolumntype{L}{>{\tiny}l} 
\newcolumntype{C}{>{\tiny}c} 
\newcolumntype{R}{>{\tiny}r} 

\newcolumntype{X}{>{\footnotesize}l} 
\newcolumntype{Z}{>{\footnotesize}c} 
\newcolumntype{D}{>{\footnotesize}r} 

\newcommand{\ie}{I\!E}
\newcommand{\iz}{I\!Z}
\newcommand{\iez}{I\!E\!Z}
\newcommand{\E}{\mathbb{E}}
\renewcommand{\d}{\mathrm{d}}
\DeclareMathOperator{\cm}{cm}

\newcommand{\defeq}{\mathrel{:=}}

\providecommand{\keywords}[1]
{
  \small	
  \emph{keywords:} #1
}

\title{Density Approximation of Affine Jump Diffusions via Closed-Form Moment Matching}
\author{Yan-Feng Wu and Jian-Qiang Hu}

\begin{document}
\maketitle
\begin{abstract}
    \noindent
    We develop a recursive approach for deriving closed-form solutions to both conditional and unconditional moments of affine jump diffusions with state-independent jump intensities. Using these moment solutions, we construct closed-form density approximations (up to a normalization constant) via moment matching for both conditional and unconditional distributions. Our framework enables important financial applications, including efficient option pricing and exact simulation for affine jump diffusions. Numerical experiments demonstrate the method's superior computational efficiency compared to existing simulation techniques, while preserving numerical precision.
\end{abstract}

\section{Introduction}
Affine jump diffusions (AJDs), introduced by \citet{duffie2000transform, duffie2003affine}, provide a powerful framework for modeling complex system dynamics while maintaining a degree of tractability. These models have found prominent applications in areas such as interest rate term structure modeling, financial asset pricing, and time series analysis. However, despite their versatility, AJDs generally lack closed-form expressions for their transition and marginal densities, even for well-known cases like the square-root diffusion \citep{cox1985theory} and the Heston stochastic volatility (SV) model \citep{heston1993closed}. Fortunately, under certain regularity conditions, the distributions of AJDs can be uniquely determined by their moments. For instance, the Heston model and its jump-extended variants are uniquely characterized by their moments \citep{kyriakou2023unified}. Moments of AJDs play a critical role in various applications, including parameter estimation \citep{bollerslev2002estimating} and exact model simulation \citep{kyriakou2023unified}. Given the importance of moments in the absence of closed-form densities, this paper investigates the explicit and closed-form computation of moments for AJDs with state-independent jump intensities---the most typical class of AJD models. Leveraging these moment solutions, we subsequently propose closed-form density approximations for the AJDs under consideration.

The literature on moments of affine processes and related processes is extensive. \citet{glasserman2010moment} analyze the tail behavior of affine diffusions, showing that their tails are either exponential or Gaussian, and characterize the range of finite moments for asset price processes derived from these diffusions.  The existence of exponential moments for affine diffusions is explored by \citet{filipovic2009affine}, while \citet{kallsen2010exponentially} demonstrate that conditional exponential moments of affine diffusions can typically be obtained by solving generalized Riccati equations. \citet{keller2015} further extend this analysis by characterizing the maximal domain of the conditional moment-generating function for affine processes. Beyond affine processes, \citet{cuchiero2012polynomial} introduce polynomial processes, a broader class of models, and show that all finite-order conditional moments are analytically tractable, up to a matrix exponential. For polynomial diffusions, which feature linear drift and quadratic diffusion terms, \citet{filipovic2016polynomial} derive closed-form expressions for conditional moments under specific conditions, with extensions to polynomial jump-diffusions provided by \citet{filipovic2020polynomial}. However, none of these works provide a method to derive both conditional and unconditional moments explicitly in closed form for the target AJDs.

In this paper, we propose a recursive method for computing both conditional and unconditional moments of affine jump diffusions with state-independent jump intensities. Our approach yields fully explicit closed-form moment formulae, facilitating fast and reliable computation of moment values without any time-consuming or error-prone numerical steps. We highlight that unconditional moments are particularly valuable for AJD models with latent states, such as stochastic volatility models (where variances are unobservable), as they characterize the stationary distributions of these processes. To support practical applications, we have developed and documented a specialized software tool for automated moment derivation \citep{wu2024ajdmom}. 

Leveraging Pearson family distributions \citep{rose2002mathstatica, kyriakou2023unified}, we construct closed-form density approximations (up to a normalization constant) for both conditional and unconditional distributions of these AJDs through moment matching. While our framework supports moment computation to arbitrary orders, numerical evidence demonstrates that matching up to the first eight moments typically achieves highly accurate density approximations. These closed-form solutions enables several important applications. 
First, the approximated densities enable efficient European option pricing through one-dimensional numerical integration of expected payoffs. Our approach thus provides a unified alternative to classical Fourier inversion methods \citep{heston1993closed, bates1996jumps, duffie2000transform} for the Heston, SVJ (SV with jumps in the return process), and SVCJ (SV with contemporaneous jumps in the return and variance processes) models. Second, the derived unconditional moments permit direct parameter estimation through method of moments for AJDs \citep{wu2023method}. Crucially, this estimation framework avoids restrictive assumptions about latent state observability---whether through option-implied proxies \citep{ai2007maximum} or high-frequency data \citep{bollerslev2002estimating}---which often prove impractical in general settings.

Our moment-matched density approximation also enable exact simulation of AJDs, offering significant computational advantages. While \citet{kyriakou2023unified} developed a moment-based simulation method for stochastic volatility models by matching integrated variance moments---yielding substantial speed improvements over Fourier inversion approaches \citep{broadie2006exact} through reduced Bessel function evaluations \citep{choudhury1996numerical}---we advance this paradigm further. Our novel scheme directly utilizes the closed-form density approximations of the models, completely eliminating Bessel function computations. This innovation proves particularly valuable for AJDs with jump-augmented latent processes like the SVCJ model, where our method outperforms existing jump-by-jump simulation schemes \citep{broadie2006exact, kyriakou2023unified}. Moreover, the availability of both conditional and unconditional moments in our framework uniquely supports equally efficient simulation from either transient or steady-state distributions.

The simulation literature for related models spans several important methodological advances. A Gamma expansion for simulating the Heston model is proposed in \citet{glasserman2011gamma}, while simulation of the SABR model is analyzed in \citet{cai2017exact} and \citet{cui2021efficient}. Exact simulation of the Ornstein–Uhlenbeck driven stochastic volatility model is explored in \citet{li2019exact}, and simulation of the Wishart multidimensional stochastic volatility model is discussed in \citet{kang2017exact}. For exact simulation of point processes with stochastic intensities, see \citet{dassios2017efficient} and \citet{giesecke2011exact}. A comprehensive overview appears in \citet{kyriakou2023unified}.

The remainder of the paper is organized as follows. Section~2 presents the recursive approach for deriving moments of the baseline AJD model, the Heston SV model. Section~3 extends the recursive method to three augmentations of the baseline model: two-factor augmentation, SV with jumps in the return, and extensions incorporating contemporaneous jumps in both the return and variance processes. Section~4 proposes density approximations via moment matching and conducts simulation experiments to validate our method. Finally, Section~5 concludes the paper. A few low-order moment formulae for the baseline AJD model are provided in the appendix for reference.

\section{A recursive approach to deriving baseline AJD moments}
The general AJD process is an n-dimensional Markov process, denoted by 
$\mathbf{x}(t)$, with state space $D\subset \mathbb{R}^n$. It is governed by the following stochastic differential equation (SDE):
$$
    \d \mathbf{x}(t) = \mathbf{\mu}(\mathbf{x}(t))\d t + \mathbf{\sigma}(\mathbf{x}(t))\d \mathbf{w}(t) + \d\mathbf{z}(t),
$$
where $\mathbf{w}(t)$ is an n-dimensional standard Wiener process, and $\mathbf{z}(t)$ is an inhomogeneous compound Poisson process (CPP). The jumps in $\mathbf{z}(t)$ occur with intensity $\mathbf{\lambda}(\mathbf{x}(t)): D\rightarrow \mathbb{R}_{\geqslant 0}^n$ and are distributed according to $F_{\mathbf{j}}(\cdot)$ on $\mathbb{R}^n$. The drift $\mathbf{\mu}(\cdot)$, instantaneous covariance matrix $\mathbf{\sigma}(\cdot)\mathbf{\sigma}(\cdot)^{\mathrm{T}}$, and jump intensity $\mathbf{\lambda}(\cdot)$ all exhibit affine dependence on the state $\mathbf{x}(t)$ \citep{duffie2000transform}, determined by coefficients $(\mathbf{K},\mathbf{H},\mathbf{l})$ as
\begin{itemize}
    \item $\mathbf{\mu}(\mathbf{x}) = \mathbf{K}_0 + \mathbf{K}_1\mathbf{x}$, for  $\mathbf{K} = (\mathbf{K}_0,\mathbf{K}_1)\in \mathbb{R}^n\times \mathbb{R}^{n\times n}$.
    \item $(\mathbf{\sigma}(\mathbf{x})\mathbf{\sigma}(\mathbf{x})^{\mathrm{T}})_{ij} = (\mathbf{H}_0)_{ij} + (\mathbf{H}_1)_{ij}\cdot \mathbf{x}$, for $\mathbf{H}=(\mathbf{H}_0,\mathbf{H}_1)\in \mathbb{R}^{n\times n}\times \mathbb{R}^{n\times n\times n}$.
    \item $\mathbf{\lambda}(\mathbf{x}) = \mathbf{l}_0 + \mathbf{l}_1\cdot \mathbf{x}$, for $\mathbf{l}=(\mathbf{l}_0,\mathbf{l}_1)\in \mathbb{R}^n\times\mathbb{R}^{n\times n}$.
\end{itemize}
%
%
%
%

AJD processes have been widely applied in financial asset valuation and other areas due to their analytical tractability and ability to capture intricate system dynamics. In this work, we focus on the most commonly used AJDs—those with state-independent jump intensities (when jumps are present). This corresponds to cases where $\mathbf{\lambda}(\mathbf{x}) = \mathbf{l}_0$. Prominent examples include the Heston SV \citep{heston1993closed}, SVJ \citep{bates1996jumps}, SVCJ \citep{duffie2000transform}, two-factor SV extensions, and other variants. We collectively refer to these as affine stochastic volatility models (ASVs). Additionally, the square-root diffusion \citep{cox1985theory} and its jump-extended versions \citep{barczy2018asymptotic, jin2016positive} also fall into this category of AJDs.

Among these AJDs, ASVs present unique challenges because part of their state variables—specifically, the volatility—is generally unobservable. As a result, the observed state variables (e.g., asset prices) do not satisfy the Markov property, making it difficult to derive the (unconditional) characteristic function for ASVs, except in a simplified Heston model \citep{jiang2002estimation}. For this reason, we primarily focus on deriving moments for ASVs, as the analysis of simpler models like the square-root diffusion with jumps is naturally embedded within the framework of the SVCJ model. The remainder of this section devotes to designing a recursive method for deriving moments of the baseline AJD model, i.e., affine stochastic volatility model without jumps, or the Heston SV model. This section lays the foundation for the extensions to other more complex AJDs.

The Heston SV model \citep{heston1993closed} serves as the baseline AJD and is described by the following system of SDEs:
\begin{align}
    \d s(t) &= \mu s(t)\d t + \sqrt{v(t)}s(t)\d w^s(t),\label{eqn:price}\\
    \d v(t) &= k(\theta - v(t))\d t + \sigma_{v}\sqrt{v(t)}\d w^v(t),\label{eqn:v-t}
\end{align}
where $s(t)$ denotes the asset price at time $t$, $\mu$ is the constant return rate, $v(t)$ is the instantaneous variance of return at time $t$, and $w^s(t)$ and $w^v(t)$ are correlated Wiener processes with correlation $\rho$. The variance process $v(t)$ follows a square-root diffusion, also known as the Cox-Ingersoll-Ross (CIR) process \citep{cox1985theory}. To ensure the positivity of the variance process $v(t)$, the Feller condition must hold: $2k\theta>\sigma_v^2$, assuming $k>0$, $\theta>0$, $\sigma_v>0$, and $v(0)>0$.  

Applying It\^{o}'s lemma \citep{shreve2004stochastic}, the dynamics of the log price $p(t)$ (where $p(t) \defeq\log s(t)$) can be expressed as:
\begin{align}
    \d p(t) = (\mu-v(t)/2)\d t + \sqrt{v(t)}\d w^s(t).\label{eqn:log-price}
\end{align}
Equations~\eqref{eqn:log-price} and \eqref{eqn:v-t} represent a specialized instance of an AJD model, with the state vector $\mathbf{x}(t) = (p(t),v(t))^{\mathrm{T}}$ and a zero arrival rate $\lambda(\mathbf{x})=0$. The return over interval $(0,t]$ is defined as $y_t \defeq p(0) - p(t)$.

The decomposition of the Wiener process $w^s(t)$ is given by:
$$
    w^s(t) = \rho w^v(t) + \sqrt{1-\rho^2}w(t),
$$
where $w(t)$ is an independent Wiener process not correlated with $w^v(t)$. 
For ease of notation, we introduce the following integral definitions:
\begin{align*}
  I_{t}&\defeq \int_0^t\sqrt{v(u)}\d w^v(u),           &I_{t}^* &\defeq \int_0^t\sqrt{v(u)}\d w(u),\\
  I\!E_{t} &\defeq \int_0^te^{ku}\sqrt{v(u)}\d w^v(u), &IV_{t} &\defeq \int_0^tv(u)\d u.
\end{align*}
The variance process $v(t)$ is a Markov process, which, as demonstrated by \citet{cox1985theory}, reaches a steady-state gamma distribution with mean $\theta$ and variance $\theta \sigma_v^2/(2k)$. We assume that $v(0)$ adheres to the steady-state distribution of $v(t)$, rendering $v(t)$ strictly stationary and ergodic as per \citet{overbeck1997}. Consequently, $p(t)$ is also stationary. Nonetheless, the results presented in the current section and next section are valid for any non-negative initial variance $v(0)$, in the steady-state context. Given the stationarity of $v(t)$ and its alignment with the steady-state gamma distribution, the moments of $v(t)$ can be expressed as follows: 
\begin{align}
    \E[v^m(t)] = \prod_{j=0}^{m-1} \left(\theta + \frac{j\sigma_v^2}{2k} \right),\label{eqn:v-moment}
\end{align}
for $m=1,2,\dots$ and $t\ge 0$.

The variable $y_t$ can be expressed by the following equation: 
$$
    y_t = \mu t - IV_t/2 + \rho I_t + \sqrt{1-\rho^2} I_t^*.
$$
The integrated variance $IV_t$ for interval $(0,t]$ is specified as:
$$
    IV_{t} = \theta (t-\tilde{t}) + \tilde{t}v_{0} -\frac{\sigma_v}{k}e^{-kt}I\!E_{t} + \frac{\sigma_v}{k}I_t,
$$
where $\tilde{t} \defeq (1-e^{-kt})/k$ and $v_0 \defeq v(0)$. It should be noted that the variance process $v(t)$ in Equation~\eqref{eqn:v-t}, pivotal to the evolution of the system, has the following solution:
$$
    v(t) = e^{-kt}v_0 + \theta\left(1-e^{-kt}\right) + \sigma_v e^{-kt}I\!E_{t}, \forall t \ge 0.
$$
Upon normalizing $y_{t}$ by its conditional expected value, we derive the centralized process $\bar{y}_{t}$:
$$
    \bar{y}_{t} \defeq y_{t} - \E[y_{t}|v_0]
$$
which can be further decomposed into its constituent terms as:
$$
    \bar{y}_{t}
    = \frac{\sigma_v}{2k} e^{-kt} I\!E_{t}
     + \left(\rho - \frac{\sigma_v}{2k} \right)I_{t} + \sqrt{1-\rho^2}I_{t}^* + \beta_{t}\theta - \beta_{t}v_{0},
$$
with $\beta_{t} \defeq (1-e^{-kt})/(2k)$.

The $m$-th central moment of $y_{t}$, denoted as $\cm_m(y_{t})$, can be determined by leveraging a set of underlying components:
\begin{align}
    \E[(e^{-kt}I\!E_{t})^{m_1}I_{t}^{m_2}(I_{t}^*)^{m_{3}}\theta^{m_4}v_{0}^{m_5}],\label{eqn:comb-moment}
\end{align}
where integers $m_i\geqslant 0$ for $i=1,2,3,4,5$ and the summation $\sum_{i=1}^{5}m_i=m$. The component moment \eqref{eqn:comb-moment} can be computed via a two-step process:
\begin{enumerate}
    \item Compute the conditional expectation by fixing $v_{0}$:
    $$
        \E[(e^{-kt}I\!E_{t})^{m_1}I_{t}^{m_2}(I_{t}^*)^{m_{3}}|v_{0}].
    $$
    \item Follow with the unconditional expectation with respect to $v_{0}$:
    $$
      \E[\E[(e^{-kt}I\!E_{t})^{m_1}I_{t}^{m_2}(I_{t}^*)^{m_{3}}|v_{0}]\theta^{m_4}v_{0}^{m_5}].
    $$
\end{enumerate}
We shall demonstrate that the conditional moment $\E[I\!E_{t}^{m_1}I_{t}^{m_2}(I_{t}^*)^{m_{3}}|v_{0}]$ can be expressed as a polynomial function of $v_{0}$. This implies that the moment described in Equation~\eqref{eqn:comb-moment} can be represented in terms of the moments of $v_{0}$. 

By using Equation~\eqref{eqn:v-moment}, we are able to calculate the moments of $v_{0}$ of any order. Consequently, this enables us to compute Equation~\eqref{eqn:comb-moment} for any given $m$.  We next analyze the expected value of the product $I\!E_{t}^{m_1}I_{t}^{m_2}(I_{t}^*)^{m_{3}}$ conditional on $v_{0}$. The term $I\!E_{t}^{m_1}I_{t}^{m_2}(I_{t}^*)^{m_{3}}$ is decomposed into two distinct components: $I\!E_{t}^{m_1}I_{t}^{m_2}$ and $(I_{t}^*)^{m_{3}}$. The decomposition is justified by the fact that these components are driven by two independent Wiener processes $w^v(t)$ and $w(t)$, respectively. 

Focusing on the differential of $I\!E_{t}^{m_1}I_{t}^{m_2}$, we express it as:
$$
    \d(I\!E_{t}^{m_1}I_{t}^{m_2})
    = c_w(t) \d w^v(t)+ c(t) \d t,
$$
where the coefficients $c_w(t)$ and $c(t)$ are given by:
$$
   c_w(t) 
   \defeq m_1 I\!E_{t}^{m_1-1}I_{t}^{m_2}\sqrt{v(t)}
    + m_2 I\!E_{t}^{m_1}I_{t}^{m_2-1}e^{kt}\sqrt{v(t)},
$$
and
$$
    c(t)
    \defeq \bigg[\frac{m_1(m_1-1)}{2}I\!E_{t}^{m_1-2}I_{t}^{m_2}e^{2kt}
     + \frac{m_2(m_2-1)}{2}I\!E_{t}^{m_1}I_{t}^{m_2-2}
     + m_1m_2I\!E_{t}^{m_1-1}I_{t}^{m_2-1}e^{kt} \bigg] v(t).
$$
For the term $I_{t}^{*m_3}$, the differential is given by:
$$
    \d I_{t}^{*m_3}
    = m_3(I_{t}^*)^{m_3-1}\sqrt{v(t)} \d w(t)
     + \frac{1}{2}m_3(m_3-1)(I_{t}^*)^{m_3-2}v(t)\d t.
$$
It is important to note that the product $\d(I\!E_{t}^{m_1}I_{t}^{m_2})\d I_{t}^{*m_3}$ is zero because the cross-variation $\d w^v(t)\d w(t)$ equals zero. Consequently, the differential of the product $I\!E_{t}^{m_1}I_{t}^{m_2}I_{t}^{*m_3}$ can be computed through: 
$$
\d(I\!E_{t}^{m_1}I_{t}^{m_2}I_{t}^{*m_3})
= (I\!E_{t}^{m_1}I_{t}^{m_2})\d I_{t}^{*m_3}
  + I_{t}^{*m_3}\d(I\!E_{t}^{m_1}I_{t}^{m_2})
$$
which produces:
\begin{align*}
  \d(I\!E_{t}^{m_1}I_{t}^{m_2}I_{t}^{*m_3})
  &=  m_3I\!E_{t}^{m_1}I_{t}^{m_2}(I_{t}^*)^{m_3-1}\sqrt{v(t)} \d w(t) 
  + c_w(t)I_{t}^{*m_3}\d w^v(t) 
  + c_3(t)\d t
\end{align*}
where
$$
c_3(t) \defeq \frac{1}{2}m_3(m_3-1) I\!E_{t}^{m_1}I_{t}^{m_2}(I_{t}^*)^{m_3-2}v(t)
  + c(t)I_{t}^{*m_3}.
$$
Discarding terms with an expected value of zero, we obtain the following conditional expectation:
$$
\E[I\!E_{t}^{m_1}I_{t}^{m_2}I_{t}^{*m_3}|v_0]
= \int_{0}^t \E[c_3(s)|v_0]ds.
$$
Furthermore, the conditional expectation $\E[I\!E_{t}^{m_1}I_{t}^{m_2}I_{t}^{*m_3}|v_0]$ can be represented in a recursive way as follows:
\begin{equation}
    \E[\ie_{t}^{m_1}I_{t}^{m_2}I_{t}^{*m_3}|v_0]
    = \sum_{j=1}^3 \left[ \frac{m_1(m_1-1)}{2}f_{3j}+ \frac{m_2(m_2-1)}{2}g_{3j}+ m_1m_2 h_{3j} + \frac{m_3(m_3-1)}{2}q_{3j} \right],\label{eqn:recursive-3items}
\end{equation}
where functions $f_{3j}, g_{3j}, h_{3j}, q_{3j}, j=1,2,3$ are defined in Table~\ref{tab:notation-in-recursive3}, with $\bar{v}_0 \defeq v_0 - \theta$. Taking $f_{31}$ as an example, it is interpreted in the following way:
\begin{equation}\label{eqn:ieii-decode}
  f_{31} \defeq \int_{0}^t e^{ks}\E[\ie_{s}^{m_1-2}I_{s}^{m_2}I_{s}^{*m_3}|v_0]\d s \cdot \bar{v}_0.
\end{equation}
The other ones are defined similarly, with reference to Table~\ref{tab:notation-in-recursive3}.
\setlength{\tabcolsep}{4pt}
\begin{table}[!ht]
    \centering
    \caption{Functions in Equation~\eqref{eqn:recursive-3items}}
    \label{tab:notation-in-recursive3}
    \begin{tabular}{XZZ ZZZ|XZZZ ZZZ} 
      \toprule
      fun  & $e^{ks}$ & $\ie_{s}$ & $I_{s}$ & $I_s^*$ & coef & fun
        & $e^{ks}$ & $\ie_{s}$ & $I_{s}$ & $I_s^*$ & coef \\
      \midrule
      $f_{31}$ & 1 & $m_1-2$ & $m_2$ & $m_3$  & $\bar{v}_0$ &
      $g_{31}$ &-1 & $m_1$   & $m_2-2$ & $m_3$ & $\bar{v}_0$\\
      $f_{32}$ & 2 & $m_1-2$ & $m_2$ & $m_3$  & $\theta$ &
      $g_{31}$ & 0 & $m_1$   & $m_2-2$ & $m_3$ & $\theta$\\
      $f_{33}$ & 1 & $m_1-1$ & $m_2$ & $m_3$  & $\sigma_v$ &
      $g_{33}$ &-1 & $m_1+1$   & $m_2-2$ & $m_3$ & $\sigma_v$\\
      $h_{31}$ & 0 & $m_1-1$ & $m_2-1$ & $m_3$  & $\bar{v}_0$ &
      $q_{31}$ &-1 & $m_1$   & $m_2$ & $m_3-2$ & $\bar{v}_0$\\
      $h_{32}$ & 1 & $m_1-1$ & $m_2-1$ & $m_3$  & $\theta$ &
      $q_{32}$ & 0 & $m_1$   & $m_2$ & $m_3-2$ & $\theta$\\
      $h_{33}$ & 0 & $m_1$ & $m_2-1$ & $m_3$  & $\sigma_v$ &
      $q_{33}$ &-1 & $m_1+1$   & $m_2$ & $m_3-2$ & $\sigma_v$\\
      \bottomrule
    \end{tabular}
\end{table}

It should be noted that $\E[I_{t}^{*m_3}|v_0] = \E[I_{t}^{m_3}|v_0]$. Utilizing Equation~\eqref{eqn:recursive-3items}, one can calculate the central moments of the variable $y_{t}$ of any desired order through a recursive procedure, which begins with the simplest combinations of $(m_1,m_2,m_3)$ where $m=1$, and progresses sequentially to more complex combinations, such as ${(m_1,m_2,m_3), m=2}$, and continues accordingly, adhering to the condition that $m_1+m_2+m_3=m$. 

The computational process, while conceptually straightforward, becomes computationally intensive and practically unfeasible for high-order moments when performed manually. To address this challenge, the recursive nature of the equations has inspired the development of a Python package, \emph{ajdmom}, which automates and streamlines the derivation process \citep{wu2024ajdmom}. This tool ensures that the application of the proposed recursive method is both efficient and accessible, even for high-order moments. To demonstrate its utility, several low-order moment formulae computed using the \emph{ajdmom} package are provided in the appendix for reference.

\section{Extensions to other affine jump diffusions}
In this section, we adapt the recursive approach introduced earlier to three extended AJD models. Adaptations to two of these AJDs are relatively straightforward. The first adaptation involves the two-factor affine SV model, which can be further generalized to three-factor, and higher-factor models using a similar methodology. The second adaptation addresses the SVJ model, which incorporates jumps in the return process. The third adaptation deals with the most complex model, the SVCJ model, where jumps occur simultaneously in both the return and variance processes. To address this complexity, we devise a new recursive method tailored to this case. Together, these extensions highlight the flexibility and broad applicability of the recursive approach in handling increasingly complex AJD settings.

\subsection{Two-factor affine stochastic volatility}
In some scenarios, researchers aim to capture two distinct streams of volatility simultaneously—one with a slow decay rate and another with a fast decay rate. This motivates the development of the two-factor stochastic volatility model, defined by the following SDEs:
\begin{align*}
    \d p(t)
    &= (\mu - v(t)/2)\d t + \sqrt{v(t)}\d w(t),\\
    v(t) &= v_1(t) + v_2(t),\\
    \d v_1(t)
    &= k_1(\theta_1-v_1(t))\d t + \sigma_{v1}\sqrt{v_1(t)}\d w_1(t),\\
    \d v_2(t)
    &= k_2(\theta_2-v_2(t))\d t + \sigma_{v2}\sqrt{v_2(t)}\d w_2(t),
\end{align*}
where $v_1(t)$ and $v_2(t)$ represent two independent CIR processes, and $w(t)$, $w_1(t)$ and $w_2(t)$ are mutually independent Wiener processes.

For simplicity of notation, we define the following new terms:
\begin{align*}
    I_{1,t} &\defeq \int_{0}^{t}\sqrt{v_1(u)}\d w_1(u),
    &I_{2,t} &\defeq \int_{0}^{t}\sqrt{v_2(u)}\d w_2(u),\\
    \ie_{1,t} &\defeq \int_{0}^{t}e^{k_1u}\sqrt{v_1(u)}\d w_1(u),\quad
    &\ie_{2,t} &\defeq \int_{0}^{t}e^{k_2u}\sqrt{v_2(u)}\d w_2(u).
\end{align*}
The term $I_{t}^*$ remains as previously defined, i.e., $I_{t}^*\equiv \int_{0}^t\sqrt{v(u)}\d w(u)$.
The centralized return, $\overline{y}_t \defeq y_t - E[y_t|v_{1,0},v_{2,0}]$, now has the following expression:
\begin{align*}
    \overline{y}_t
    &= \frac{\sigma_{v1}}{2k_1}e^{-k_1 t}I\!E_{1,t}
     - \frac{\sigma_{v1}}{2k_1}I_{1,t}
     + \frac{\sigma_{v2}}{2k_2}e^{-k_2 t}I\!E_{2,t}
     - \frac{\sigma_{v2}}{2k_2}I_{2,t}
     + I_{t}^{*}\\
    &\quad + \frac{1}{2}(\tilde{t}_1\theta_1 + \tilde{t}_2\theta_2)
    - \frac{1}{2}\tilde{t}_1v_{1,0} - \frac{1}{2}\tilde{t}_2v_{2,0},
\end{align*}
where $\tilde{t}_i \defeq (1-e^{-k_it})/k_i, i=1,2$.
Thus, the essential computation in deriving moments of the centralized return $\bar{y}_t$ reduces to the computation of the joint conditional moments: 
\begin{equation}\label{eqn:m1m2m3m4m5}
  \E[\ie_{1,t}^{m_1}I_{1,t}^{m_2}\ie_{2,t}^{m_3}I_{2,t}^{m_4}I_{t}^{*m_5}|v_{1,0}, v_{2,0}].
\end{equation}

The recursive equation for computing the joint conditional moments in Equation~\eqref{eqn:m1m2m3m4m5} takes the following form:
\begin{align}
    &\E[\ie_{1,t}^{m_1}I_{1,t}^{m_2}\ie_{2,t}^{m_3}I_{2,t}^{m_4}I_{t}^{*m_5}|v_{1,0}, v_{2,0}]
    =\sum_{j=1}^3 \left[ \frac{m_1(m_1-1)}{2}f_{5j}^1+ \frac{m_2(m_2-1)}{2}g_{5j}^1 + m_1m_2h_{5j}^1 \right]\nonumber\\
    &\qquad +\sum_{j=1}^3 \left[ \frac{m_3(m_3-1)}{2} f_{5j}^2 + \frac{m_4(m_4-1)}{2} g_{5j}^2+ m_3m_4 h_{5j}^2 \right] + \sum_{j=1}^3 \left[ \frac{m_5(m_5-1)}{2}(q_{5j}^1 + q_{5j}^2)\right], \label{eqn:recursive-iei-iei-i}
\end{align}
where $f_{5j}^i, g_{5j}^i, h_{5j}^i, q_{5j}^i$, $i=1,2$, $j=1,2,3$, are defined in Table~\ref{tab:notation}, with $\bar{v}_{1,0} \defeq v_{1,0} - \theta_1$ and $\bar{v}_{2,0} \defeq v_{2,0} - \theta_2$. 
For example, $f_{51}^1$ is defined as:
\begin{equation}\label{eqn:iei-iei-i-decode}
    f_{51}^1 \defeq \int_0^t e^{k_1s} \E[\ie_{1,s}^{m_1}I_{1,s}^{m_2-2}\ie_{2,s}^{m_3}I_{2,s}^{m_4}I_{s}^{*m_5}|v_{1,0},v_{2,0}]\d s \cdot \bar{v}_{1,0}.
\end{equation}
The other functions in Equation~\eqref{eqn:recursive-iei-iei-i} are defined in a similar manner, as detailed in Table~\ref{tab:notation}.
\setlength{\tabcolsep}{3pt}
\begin{table}[!ht]
    \centering
    \caption{Functions in Equation~\eqref{eqn:recursive-iei-iei-i}}
    \label{tab:notation}
    \begin{tabular}{XZZZ ZZZZ|XZZZ ZZZZ} 
      \toprule
      fun  & $e^{k_1s}$ & $\ie_{1,s}$ & $I_{1,s}$ & $\ie_{2,s}$ & $I_{2,s}$ & $I_s^*$ & coef & fun
        & $e^{k_2s}$ & $\ie_{1,s}$ & $I_{1,s}$ & $\ie_{2,s}$ & $I_{2,s}$ & $I_s^*$ & coef \\
      \midrule
      $f_{51}^1$ & 1 & $m_1-2$ & $m_2$ & $m_3$ & $m_4$ & $m_5$ & $\bar{v}_{1,0}$ & $f_{51}^2$ & 1 & $m_1$ & $m_2$ & $m_3-1$ & $m_4$ & $m_5$ & $\bar{v}_{2,0}$ \\
      $f_{52}^1$ & 2 & $m_1-2$ & $m_2$ & $m_3$ & $m_4$ & $m_5$ & $\theta_1$ & $f_{52}^2$ & 2 & $m_1$ & $m_2$ & $m_3-1$ & $m_4$ & $m_5$ & $\theta_2$ \\
      $f_{53}^1$ & 1 & $m_1-1$ & $m_2$ & $m_3$ & $m_4$ & $m_5$ & $\sigma_{v1}$ & $f_{53}^2$ & 1 & $m_1$ & $m_2$ & $m_3-1$ & $m_4$ & $m_5$ & $\sigma_{v2}$ \\
      $g_{51}^1$ &-1 & $m_1$ & $m_2-2$ & $m_3$ & $m_4$ & $m_5$ & $\bar{v}_{1,0}$ & $g_{51}^2$ &-1 & $m_1$ & $m_2$ & $m_3$ & $m_4-2$ & $m_5$ & $\bar{v}_{2,0}$ \\
      $g_{52}^1$ & 0 & $m_1$ & $m_2-2$ & $m_3$ & $m_4$ & $m_5$ & $\theta_1$ & $g_{52}^2$ & 0 & $m_1$ & $m_2$ & $m_3$ & $m_4-2$ & $m_5$ & $\theta_2$ \\
      $g_{53}^1$ &-1 & $m_1+1$ & $m_2-2$ & $m_3$ & $m_4$ & $m_5$ & $\sigma_{v1}$ & $g_{53}^2$ &-1 & $m_1$ & $m_2$ & $m_3+1$ & $m_4-2$ & $m_5$ & $\sigma_{v2}$ \\
      $h_{51}^1$ & 0 & $m_1-1$ & $m_2-1$ & $m_3$ & $m_4$ & $m_5$ & $\bar{v}_{1,0}$ & $h_{51}^2$ & 0 & $m_1$ & $m_2$ & $m_3-1$ & $m_4-1$ & $m_5$ & $\bar{v}_{2,0}$ \\
      $h_{52}^1$ & 1 & $m_1-1$ & $m_2-1$ & $m_3$ & $m_4$ & $m_5$ & $\theta_1$ & $h_{52}^2$ & 1 & $m_1$ & $m_2$ & $m_3-1$ & $m_4-1$ & $m_5$ & $\theta_2$ \\
      $h_{53}^1$ & 0 & $m_1$ & $m_2-1$ & $m_3$ & $m_4$ & $m_5$ & $\sigma_{v1}$ & $h_{53}^2$ & 0 & $m_1$ & $m_2$ & $m_3$ & $m_4-1$ & $m_5$ & $\sigma_{v2}$ \\
      $q_{51}^1$ &-1 & $m_1$ & $m_2$ & $m_3$ & $m_4$ & $m_5-2$ & $\bar{v}_{1,0}$ & $q_{51}^2$ &-1 & $m_1$ & $m_2$ & $m_3$ & $m_4$ & $m_5-2$ & $\bar{v}_{2,0}$ \\
      $q_{52}^1$ & 0 & $m_1$ & $m_2$ & $m_3$ & $m_4$ & $m_5-2$ & $\theta_1$ & $q_{52}^2$ & 0 & $m_1$ & $m_2$ & $m_3$ & $m_4$ & $m_5-2$ & $\theta_2$ \\
      $q_{53}^1$ &-1 & $m_1+1$ & $m_2$ & $m_3$ & $m_4$ & $m_5-2$ & $\sigma_{v1}$ & $q_{53}^2$ &-1 & $m_1$ & $m_2$ & $m_3+1$ & $m_4$ & $m_5-2$ & $\sigma_{v2}$ \\
      \bottomrule
    \end{tabular}
\end{table}

\subsection{Affine stochastic volatility with jumps in returns}
The stochastic volatility with jumps incorporated into the return process, is described by the following SDEs:
\begin{align*}
    \d p(t)
    &= (\mu - v(t)/2)\d t + \sqrt{v(t)}\d w^s(t) + \d z(t),\\
    \d v(t)
    &= k(\theta-v(t))\d t + \sigma\sqrt{v(t)}\d w^v(t),
\end{align*}
where $z(t)$ represents a CPP with a constant arrival rate $\lambda$ and jump size distribution $F_j(\cdot,\mathbf{\theta}_j)$ parameterized by $\mathbf{\theta}_j$. The remaining parameters are consistent with those in the Heston SV model as delineated by Equations~\eqref{eqn:log-price} and \eqref{eqn:v-t}. 

The moments for this model follow an analogous derivation process as for the Heston SV model. Here, we decompose $y_t$ as:
$$
    y_t = y_{o,t} + J_t,
$$
where $y_{o,t}$ denotes the return of the Heston SV model, i.e.,
$$
    y_{o,t} \equiv \mu t - \frac{1}{2}IV_{t} + \rho I_t +
\sqrt{1-\rho^2}I_t^{*},
$$
and
$$
    J_t \equiv z(t) - z(0) = \sum_{i=N(0)+1}^{N(t)}j_i,
$$
$N(t)$ denotes the Poisson process associated with the CPP $z(t)$, and $j_i$ follows a normal distribution $\mathcal{N}(\mu_j,\sigma_j^2)$. Consequently, the $m$-th moment of $y_t$ can be computed as follows:
\begin{equation}\label{eqn:svj-moment}
    \E[y_t^m] = \E[(y_{o,t}+J_t)^m] =  \sum_{i=0}^m \binom{m}{i} \E[y_{o,t}^i]\E[J_t^{m-i}],
\end{equation}
where $\binom{m}{i}$ represents the binomial coefficient, the number of ways to choose $i$ elements from a set of $m$ distinct elements. Therefore, we can easily compute moments of the affine stochastic volatility with jumps in the return model via Equation~\eqref{eqn:svj-moment} once moments of the Heston model are computed. 

\subsection{Affine stochastic volatility with contemporaneous jumps}
In this subsection, we address the most sophisticated affine stochastic volatility model, the SVCJ model. We first present a new recursive method for computing conditional moments of the SVCJ model, conditioned on the initial variance. It will be shown that these conditional moment formulae are polynomials in the initial variance. Next, we introduce another recursive approach, detailed in the appendix, to compute the unconditional moments of the variance. These moments can then be used to derive the unconditional moment formulae for the SVCJ model.

The SVCJ model augments the Heston model by incorporating contemporaneous jumps into the return and variance, described by the following SDEs:
\begin{align}
    \d p(t) &= (\mu- v(t)/2) \d t + \sqrt{v(t)}\d w^s(t) + \d z^s(t),\label{eqn:svcj-price}\\
    \d v(t) &= k(\theta - v(t))\d t + \sigma_v \sqrt{v(t)}\d w^v(t) + \d z^v(t),\label{eqn:svcj-variance}
\end{align}
where $z^v(t)$ is a compound Poisson process with constant arrival rate $\lambda$ and jumps ($J_i^v$) distributed according to an exponential distribution with scale parameter $\mu_v$, 
$z^s(t)$ is another compound Poisson process sharing the same arrival process with $z^v(t)$, and with jumps distributed according to a normal distribution with mean $\mu_s + \rho_JJ_i^v$ and variance $\sigma_s^2$, all other settings kept as the same in the Heston model.

We adopt the same definitions for $\ie_t$, $I_t$ and $I_t^*$ as in the Heston model.
Additionally, we introduce the following notation:
\begin{equation*}
    \iez_t \defeq \int_0^t e^{ks}\d z^v(s), ~
    \iz_t \defeq \int_0^t \d z^v(s), ~
    \iz_t^s \defeq \int_0^t \d z^s(t), ~
    \iz_t^* \defeq \int_0^t\d z^*(t),
\end{equation*}
where $z^*(t)$ is another compound Poisson process sharing the same arrival process as $z^v(t)$ and $z^s(t)$, but with independent jumps $J_i^*\sim \mathcal{N}(\mu_s, \sigma_s^2)$.
Since the jumps of $z^s(t)$ are distributed as $J_i^s|J_i^v \sim \mathcal{N}(\mu_s + \rho_J J_i^v, \sigma_s^2)$, the compound Poisson process in the return can be decomposed into another two compound Poisson processes, i.e., $\iz_{t}^s = \rho_J \iz_t + \iz_t^*$. The solution to the variance process in Equation~\eqref{eqn:svcj-variance} now is given by:
\begin{equation}\label{eqn:svcj-variance-solution1}
    e^{kt}v_t = (v_0-\theta) + e^{kt}\theta + \sigma_v \ie_t + \iez_t.
\end{equation}
For the return $y_t$ ($y_t \equiv p(t) - p(0)$), we have the following decomposition:
\begin{align*}
    y_t
    &= \frac{\sigma_v}{2k} e^{-kt}\ie_t +
     \left(\rho -\frac{\sigma_v}{2k} \right)I_t + \sqrt{1-\rho^2}I_t^{*}
     + \frac{1}{2k}e^{-kt}\iez_t + \left(\rho_J - \frac{1}{2k}\right)\iz_t + \iz_t^{*}\\
    &\quad + \left(\mu-\frac{\theta}{2}\right)t - (v_0 - \theta)\beta_t,
\end{align*}
where $\beta_t = (1-e^{-kt})/(2k)$. 
The first conditional moment is straightforward to compute and is given by:
\begin{equation*}
    \E[y_t|v_0] 
    = (\mu - \E[v]/2)t - (v_0 - \E[v])\beta_{t} + \lambda t (\mu_s + \rho_J\mu_v),
\end{equation*}
where $\E[v] = \theta + \lambda \mu_v /k$.
To compute higher-order conditional moments of $y_t$, it suffices to evaluate the conditional joint moment:
\begin{equation}\label{eqn:joint-ieii-ieziziz}
    \E[\ie_t^{m_1}I_t^{m_2}I_t^{*m_3}\iez_t^{m_4}\iz_t^{m_5}\iz_t^{*m_6}|v_0].
\end{equation}
For cases where $m_1+m_2=1$ or $m_3 = 1$, the conditional joint moment in Equation~\eqref{eqn:joint-ieii-ieziziz} evaluates to 0. Hereafter, we will consider the typical cases that does not evaluate to 0. Before addressing the computation of this conditional joint moment, we outline the derivation of conditional central moment of $y_t$. Define the centralized return as
\begin{equation*}
    \bar{y}_t \defeq y_t - \E[y_t|v_0].
\end{equation*}
The $m$-th conditional central moment of $y_t$ can then be expressed as:
\begin{align*}
    \E[\bar{y}_t^m|v_0] = \sum_{i=0}^m\binom{m}{i}(-1)^i \E^i[y_t|v_0]\E[y_t^{m-i}|v_0].
\end{align*}
This decomposition demonstrates that the computation of conditional central moments relies on the computation of conditional moments. 

We now shift our focus to computing the conditional joint moment in Equation~\eqref{eqn:joint-ieii-ieziziz}. While the recursive computation of $\E[\ie_t^{m_1} I_t^{m_2} I_t^{*m_3}|v_0]$ is well-established for models such as the Heston model, we encounter a new challenge: the last three quantities $\iez_t^{m_4} \iz_t^{m_5} \iz_t^{*m_6}$ are not independent of the first three quantities $\ie_t^{m_1} I_t^{m_2} I_t^{*m_3}$ in Equation~\eqref{eqn:joint-ieii-ieziziz}. For $m_4 + m_5 + m_6 \ge 1$, we must evaluate integrals of the form
\begin{equation}\label{eqn:sample-term}
    \int_0^t e^{lks} \E[\ie_s^{m_1}I_s^{m_2}I_s^{*m_3} \iez_t^{m_4} \iz_t^{m_5} \iz_t^{*m_6}|v_0]\d s,
\end{equation}
where $l$ is an integer number.
The dependence between $\ie_s^{m_1}I_s^{m_2}I_s^{*m_3}$ and $\iez_t^{m_4} \iz_t^{m_5} \iz_t^{*m_6}$ motivates us to decompose the latter as follows:
\begin{align*}
    &\iez_t^{m_4}\iz_t^{m_5}\iz_t^{*m_6}\\
    &= \sum_{i_1=0}^{m_4}\sum_{i_2=0}^{m_5}\sum_{i_3=0}^{m_6} \binom{m_4}{i_1}\binom{m_5}{i_2}\binom{m_6}{i_3} \iez_s^{i_1}\iz_s^{i_2}\iz_s^{*i_3} \iez_{s,t}^{m_4-i_1} \iz_{s,t}^{m_5-i_2} \iz_{s,t}^{*m_6-i_3}, \quad \forall s \le t,
\end{align*}
where $\iez_t$ is split into two independent parts $\iez_s, \iez_{s,t}$, i.e., $\iez_t = \iez_s + \iez_{s,t}$. Similarly, $\iz_t$ and $\iz_t^*$ are decomposed as $\iz_t = \iz_s + \iz_{s,t}$ and $\iz_t^* = \iz_s^* + \iz_{s,t}^*$, respectively. Here, the new terms $\iez_{s,t}$, $\iz_{s,t}$ and $\iz_{s,t}^{*}$ are defined as
\begin{equation*}
    \iez_{s,t} \defeq \int_s^te^{ku}\d z^v(u), \quad \iz_{s,t} \defeq \int_s^t\d z^v(u), \quad \iz^{*}_{s,t} \defeq \int_s^t\d z^{s}(u).
\end{equation*}
Consequently, Equation~\eqref{eqn:sample-term} can be evaluated as
\begin{align*}
    &\int_0^t e^{lks} \E[\ie_s^{m_1}I_s^{m_2}I_s^{*m_3} \iez_t^{m_4} \iz_t^{m_5} \iz_t^{*m_6}|v_0]\d s\\
    &= \sum_{i_1=0}^{m_4}\sum_{i_2=0}^{m_5}\sum_{i_3=0}^{m_6}\binom{m_4}{i_1}\binom{m_5}{i_2}\binom{m_6}{i_3} \int_0^t e^{lks} \E[\ie_s^{m_1}I_s^{m_2}I_s^{*m_3} \iez_s^{i_1}\iz_s^{i_2}\iz_s^{*i_3}|v_0] \cdot M(i_1,i_2,i_3)\d s.
\end{align*}
where $M(i_1,i_2,i_3)\defeq \E[\iez_{s,t}^{m_4-i_1} \iz_{s,t}^{m_5-i_2} \iz_{s,t}^{*m_6-i_3}|v_0]$. We have established that the conditional joint moment $\E[\iez_{s,t}^{m_4}\iz_{s,t}^{m_5}\iz_{s,t}^{*m_6}|v_0]$ can be computed as the following ``polynomial":
\begin{equation}\label{eqn:ieziziz-poly}
    \E[\iez_{s,t}^{m_4}\iz_{s,t}^{m_5}\iz_{s,t}^{*m_6}|v_0]
    = \sum_{\mathbf{j}}c_{\mathbf{j}}e^{j_1kt}t^{j_2}e^{j_3ks}s^{j_4}k^{-j_5}\lambda^{j_6}\mu_v^{j_7}\mu_s^{j_8}\sigma_s^{j_9},
\end{equation}
where $\mathbf{j}\defeq (j_1,\dots, j_9)$, $j_1,\dots,j_9$ are integers, $c_{\mathbf{j}}$ represents the corresponding monomial coefficient. For detailed derivations, please refer to Appendix~B.

With Equation~\eqref{eqn:ieziziz-poly}, the conditional joint moment in Equation~\eqref{eqn:joint-ieii-ieziziz} can be computed recursively as follows: for $m_1+m_2+m_3\ge 2$, $m_3\neq 1$ and $m_i \ge 0, i=1,\dots,6$,
\begin{align}
    &\E[\ie_t^{m_1} I_t^{m_2} I_t^{*m_3} \iez_t^{m_4} \iz_t^{m_5} \iz_t^{*m_6}|v_0]\nonumber\\
    &=\sum_{i_1=0}^{m_4}\sum_{i_2=0}^{m_5}\sum_{i_3=0}^{m_6}\binom{m_4}{i_1}\binom{m_5}{i_2}\binom{m_6}{i_3}\sum_{\mathbf{j}}c_{\mathbf{j}}e^{j_1kt}t^{j_2}k^{-j_5}\lambda^{j_6}\mu_v^{j_7}\mu_s^{j_8}\sigma_s^{j_9}F(m_1, m_2, m_3),\label{eqn:recursive-ieii-ieziziz}
\end{align}
where 
\begin{equation}\label{eqn:F-m1-m2-m3}
    F(m_1,m_2,m_3) 
    \defeq \sum_{j=1}^4\left[\frac{m_1(m_1-1)}{2}f_{6j}
      + \frac{m_2(m_2-1)}{2}g_{6j}
      + m_1m_2h_{6j}
      + \frac{m_3(m_3-1)}{2}q_{6j}\right],
\end{equation}
and functions $f_{6j}, g_{6j}, h_{6j}, q_{6j}, j=1,2,3,4$ are defined in Table~\ref{tab:notation-in-recursive-ieii-ieziziz}. For instance, $f_{61}$ is defined as 
\begin{equation}\label{eqn:ieii-ieziziz-decode}
 f_{61} \defeq \int_0^te^{(j_3+1)ks}s^{j_4}\E[\ie_s^{m_1-2}I_s^{m_2}I_s^{*m_3}\iez_s^{i_1}\iz_s^{i_2}\iz_s^{*i_3}|v_0]\d s \times (v_0-\theta).
\end{equation}
The other functions are defined in a similar way, as detailed in Table~\ref{tab:notation-in-recursive-ieii-ieziziz}.
\setlength{\tabcolsep}{8pt}
\begin{table}[!ht]
    \centering
    \caption{Functions in Equation~\eqref{eqn:F-m1-m2-m3}}
    \label{tab:notation-in-recursive-ieii-ieziziz}
    \begin{tabular}{XZZZZ ZZZZZ} 
      \toprule
      fun  & $e^{ks}$ & $s$  & $\ie_{s}$ & $I_{s}$ & $I_s^*$ & $\iez_s$ & $\iz_s$ & $\iz_s^*$ & coef\\
      \midrule
      $f_{61}$ & $j_3+1$ & $j_4$ & $m_1-2$ & $m_2$ & $m_3$ & $i_1$ & $i_2$ & $i_3$ & $v_0 - \theta$\\
      $f_{62}$ & $j_3+2$ & $j_4$ & $m_1-2$ & $m_2$ & $m_3$ & $i_1$ & $i_2$ & $i_3$ & $\theta$\\
      $f_{63}$ & $j_3+1$ & $j_4$ & $m_1-1$ & $m_2$ & $m_3$ & $i_1$ & $i_2$ & $i_3$ & $\sigma_v$\\
      $f_{64}$ & $j_3+1$ & $j_4$ & $m_1-2$ & $m_2$ & $m_3$ & $i_1+1$ & $i_2$ & $i_3$ & 1\\
      $g_{61}$ & $j_3-1$ & $j_4$ & $m_1$ & $m_2-2$ & $m_3$ & $i_1$ & $i_2$ & $i_3$ & $v_0 - \theta$\\
      $g_{62}$ & $j_3$ & $j_4$ & $m_1$ & $m_2-2$ & $m_3$ & $i_1$ & $i_2$ & $i_3$ & $\theta$\\
      $g_{63}$ & $j_3-1$ & $j_4$ & $m_1+1$ & $m_2-2$ & $m_3$ & $i_1$ & $i_2$ & $i_3$ & $\sigma_v$\\
      $g_{64}$ & $j_3-1$ & $j_4$ & $m_1$ & $m_2-2$ & $m_3$ & $i_1+1$ & $i_2$ & $i_3$ & 1\\
      $h_{61}$ & $j_3$ & $j_4$ & $m_1-1$ & $m_2-1$ & $m_3$ & $i_1$ & $i_2$ & $i_3$ & $v_0 - \theta$\\
      $h_{62}$ & $j_3+1$ & $j_4$ & $m_1-1$ & $m_2-1$ & $m_3$ & $i_1$ & $i_2$ & $i_3$ & $\theta$\\
      $h_{63}$ & $j_3$ & $j_4$ & $m_1$ & $m_2-1$ & $m_3$ & $i_1$ & $i_2$ & $i_3$ & $\sigma_v$\\
      $h_{64}$ & $j_3$ & $j_4$ & $m_1-1$ & $m_2-1$ & $m_3$ & $i_1+1$ & $i_2$ & $i_3$ & 1\\
      $q_{61}$ & $j_3-1$ & $j_4$ & $m_1$ & $m_2$ & $m_3-2$ & $i_1$ & $i_2$ & $i_3$ & $v_0 - \theta$\\
      $q_{62}$ & $j_3$ & $j_4$ & $m_1$ & $m_2$ & $m_3-2$ & $i_1$ & $i_2$ & $i_3$ & $\theta$\\
      $q_{63}$ & $j_3-1$ & $j_4$ & $m_1+1$ & $m_2$ & $m_3-2$ & $i_1$ & $i_2$ & $i_3$ & $\sigma_v$\\
      $q_{64}$ & $j_3-1$ & $j_4$ & $m_1$ & $m_2$ & $m_3-2$ & $i_1+1$ & $i_2$ & $i_3$ & 1\\
      \bottomrule
    \end{tabular}
\end{table}

Before closing this subsection, we highlight that the final expression for the conditional joint moment in Equation~\eqref{eqn:joint-ieii-ieziziz} takes the form of a polynomial in $v_0-\theta$. Specifically, it can be expressed as:
\begin{equation}\label{eqn:ieii-ieziziz-polynomial}
    \E[\ie_t^{m_1}I_t^{m_2}I_t^{*m_3}\iez_t^{m_4}\iz_t^{m_5}\iz_t^{*m_6}|v_0]
    = \sum_{i=0}^{\lfloor (m_1+m_2)/2 \rfloor + \lfloor m_3/2 \rfloor} c_i (v_0-\theta)^i,
\end{equation}
where $\lfloor x \rfloor$ denotes the floor function, i.e., the greatest integer less than or equal to $x$. Here, with a slight abuse of notation, $c_i$ represents the coefficient, which can be computed via the recursive Equation~\eqref{eqn:recursive-ieii-ieziziz}. Consequently, the conditional moment of the return, $\E[y_t^m|v_0]$, is also a polynomial in $v_0$. This property allows us to leverage the polynomial structure to compute the unconditional moments of the return, $\E[y_t^m]$, as demonstrated in Appendix~C.

\section{Numerical experiments}
Building on our moment solutions, this section presents a general moment-matched density approximation framework applicable to all AJD models. We validate the accuracy and computational efficiency of these approximations through extensive simulation experiments, evaluating both conditional distributions (Heston, SVJ, and SVCJ models) and the unconditional distribution of the SVCJ model.

\subsection{Moment-matched density approximation}
Given moments derived by our method, the unknown true densities of the AJDs can be approximated accurately via the generalized Pearson family of distributions \citep{rose2002mathstatica, kyriakou2023unified}. Let us denote the unknown true density and the Pearson density by $p(x)$ and $\tilde{p}(x)$, respectively. The Pearson density is described by the following differential equation: 
\begin{equation}\label{eqn:pearson-density-ode}
    \frac{\d \tilde{p}(x)}{\d x} = -\frac{a+x}{c_0 + c_1x + \cdots + c_nx^n} \tilde{p}(x),\quad n \ge 2,
\end{equation}
where coefficients $a, c_0, \dots, c_n$ are determined from moments of the true distribution. Under the classical Pearson setting, $n=2$, and we can match the first four moments of  $\tilde{p}(x)$ with those of $p(x)$. For $n=3$, match the first six moments; $n=4$, match the first eight moments, so on and so forth. The unknown true density $p(x)$ can be accurately approximated by the Pearson density $\tilde{p}(x)$ in the asymptotic way.

We next show how to recover the coefficients $a,c_0,\dots,c_n$ from moments of the target distribution. The moment notation, $\tilde{\mu}_m \defeq \int_{-\infty}^{\infty}x^m\tilde{p}(x)dx$, is introduced for simplicity. With moderate regulations on $\tilde{p}(x)$ at the extreme points \citep{rose2002mathstatica}, we have the following moment recurrent equation:
\begin{equation*}
    c_0 m \tilde{\mu}_{m-1} + c_1 (m+1) \tilde{\mu}_m + \cdots + c_n (n+m)\tilde{\mu}_{n+m-1} = a\tilde{\mu}_m + \tilde{\mu}_{m+1}.
\end{equation*}
Since there are $n+2$ unknown coefficients, i.e., $a, c_0,\dots, c_n$, we need $n+2$ equations. Let $m = 0, 1, \dots, n+1$, we get the following system of linear equations:
\begin{equation}\label{eqn:moms-to-coefs}
    \left[
    \begin{matrix}
      -1 & 0 & 1 & \cdots & n\tilde{\mu}_{n-1}\\
      -\tilde{\mu}_1 & 1 & 2\tilde{\mu}_1 & \cdots & (n+1)\tilde{\mu}_{n}\\
      -\tilde{\mu}_2 & 2\tilde{\mu}_1 & 3\tilde{\mu}_2 & \cdots & (n+2)\tilde{\mu}_{n+1}\\
      \vdots & \vdots & \vdots & \ddots & \vdots \\
      -\tilde{\mu}_{n+1} & (n+1)\tilde{\mu}_{n} & (n+2)\tilde{\mu}_{n+1} & \cdots & (2n+1)\tilde{\mu}_{2n}
    \end{matrix}
    \right]
    \left[
    \begin{matrix}
        a\\ c_0 \\ c_1 \\ \vdots \\ c_n
    \end{matrix}
    \right]
    = 
    \left[
    \begin{matrix}
        \tilde{\mu}_1 \\ \tilde{\mu}_2 \\ \tilde{\mu}_3 \\ \vdots \\ \tilde{\mu}_{n+2}
    \end{matrix}
    \right].
\end{equation}
The system of equations \eqref{eqn:moms-to-coefs} consumes $2n$ moments, i.e., $\tilde{\mu}_1, \cdots, \tilde{\mu}_{2n}$. Given these $2n$ moments, it is easy to solve the system of linear equations to get values of the coefficients.
Define the $m$-th moment of the unknown true distribution $p(x)$ as 
$\mu_m \defeq \int_{-\infty}^{\infty}x^m p(x)dx$. Letting $\tilde{\mu}_i = \mu_i, i=1,\cdots,2n$, we get an approximation $\tilde{p}(x)$ of $p(x)$.

We then demonstrate that the density $\tilde{p}(x)$ solution to Equation~\eqref{eqn:pearson-density-ode} takes a closed-form expression, up to a normalization constant. It is easy to have from Equation~\eqref{eqn:pearson-density-ode}
\begin{equation}\label{eqn:pearson-density}
    \tilde{p}(x) \propto \mathrm{exp}\left(-\int \frac{a+x}{c_0 + c_1x + \cdots + c_nx^n}\d x\right).
\end{equation}
By partial fraction decomposition, the rational function
$$
 \frac{a+x}{c_0+c_1x+\cdots+c_nx^n}
$$
can be split into a series of simpler fractions with denominators of order 1 or 2. Thus makes the integral of the rational function can be solved explicitly, yielding an explicit and closed-form expression for the right hand side of Equation~\eqref{eqn:pearson-density}. Let us introduce notation $\tilde{p}_{0}(x)$ to denote the un-normarlized density:
\begin{equation*}
    \tilde{p}_{0}(x) \defeq \mathrm{exp}\left(-\int \frac{a+x}{c_0 + c_1x + \cdots + c_nx^n}\d x\right).
\end{equation*}
To evaluate the density function $\tilde{p}(x)$, we need to resort to numerical integration only once to get the normalizing constant
\begin{equation*}
    C \defeq \int_{-\infty}^{\infty} \tilde{p}_0(x) \d x.
\end{equation*}
During the numerical integration, the support of $\tilde{p}_0(x)$ can be truncated to be $[\mu_1 - l \sigma, \mu_1 + u\sigma]$ where $\sigma$ denotes the standard deviation of $p(x)$, $l$ and $u$ are used to control lower and upper bounds of the support, respectively. Usually, $l$ and $u$ can be set as 7, fine tune of which can be achieved according to the skewness and kurtosis of $p(x)$. For instance, a negative skewness means a longer left tail, a typical case for the SV models considered in finance. A more efficient way is to determine the support according to quantiles of the specific classic Pearson distribution.

\begin{figure}[!ht]
    \centering
    \includegraphics[width=1\linewidth]{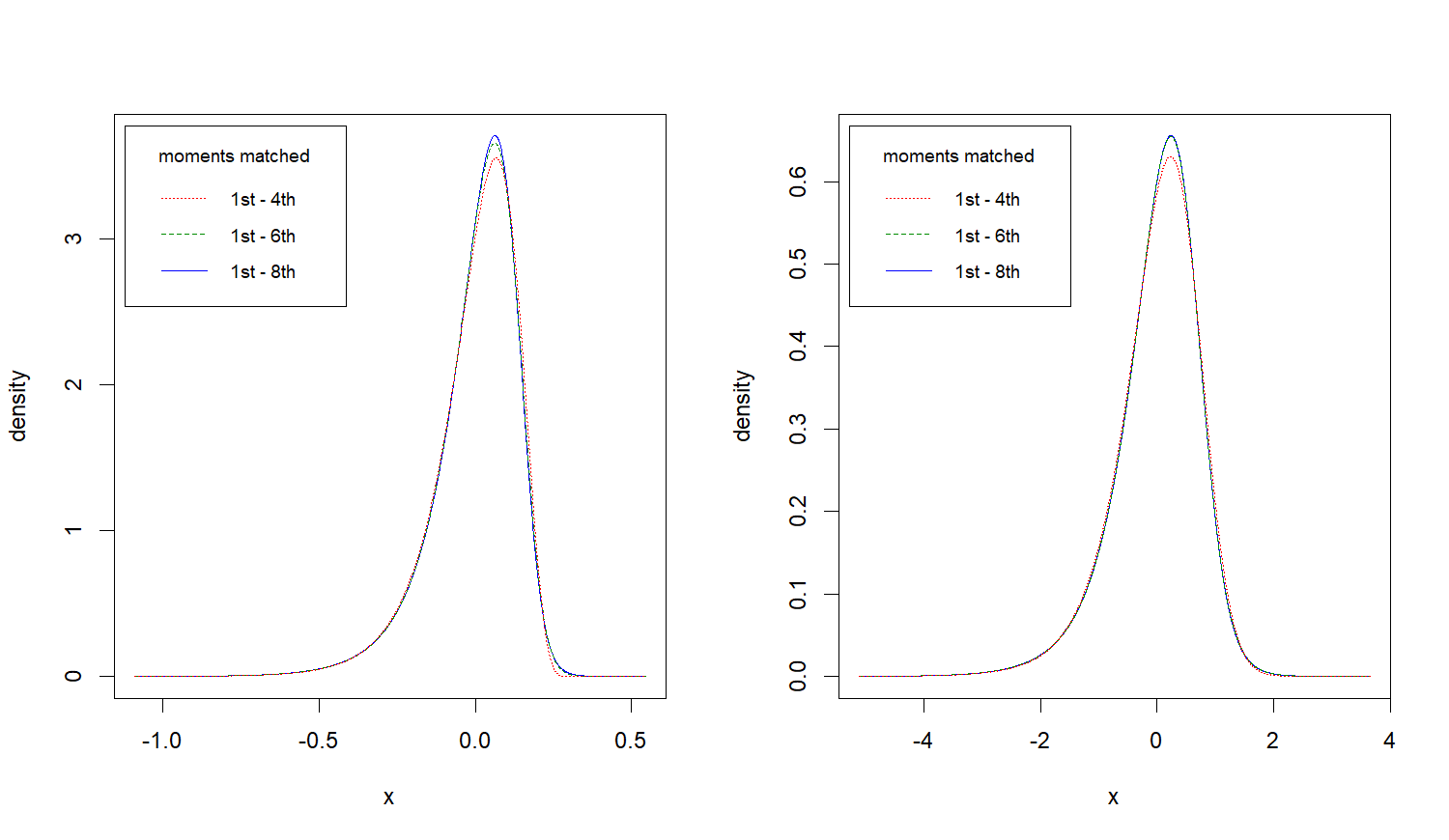}
    \caption{Density approximations by matching moments in Table~\ref{tab:example-moments}}
    \label{fig:density-approximation}
\end{figure}

We first conduct experiments to demonstrate that the convergence of the approximated densities can usually be achieved by matching at most the first eight moments. Table~\ref{tab:example-moments} presents two groups of central moments, which are computed from the derived moment formulae of the Heston model, with parameters set as in Tables~\ref{tab:heston-1} and \ref{tab:heston-2}, respectively. The first row corresponds to a case that the central moment absolute values decrease along with the order of the moments. While, the second row represents another case that the central moment absolute values increase along with the order of the moments, except for the third central moment $\mu_3$. We match the moments sequentially, i.e., first match the first four moments, then match the first six moments and last match the first eight moments. We plot the density sequential approximations together to see their convergence behavior, as shown in Figure~\ref{fig:density-approximation}. It shows that matching the first eight moments is usually enough to achieve the convergence of the density approximation, see \citet{kyriakou2023unified} for a theoretical analysis.

\begin{table}[!ht]
    \centering
    \caption{Central moment examples}
    \setlength{\tabcolsep}{5pt}
    \label{tab:example-moments}
    \begin{tabular}{X DDDD DDDD}
      \toprule
       & $\bar{\mu}_1$ & $\bar{\mu}_2$ & $\bar{\mu}_3$ & $\bar{\mu}_4$ & $\bar{\mu}_5$ & $\bar{\mu}_6$ & $\bar{\mu}_7$ & $\bar{\mu}_8$ \\ 
       \midrule
       case 1 & 0 & 0.0186180 &-0.0033707 &0.0022472 &-0.001222 &0.0009233 &-0.0007855 &0.0007777\\
       case 2 & 0 &0.5346568 &-0.3868392 &1.6363802 &-3.994397 &16.5271025 &-70.1374506 &365.1618607\\
       \bottomrule
    \end{tabular}
\end{table}

Given the probability density, there are plenty of algorithms to generate samples from the distribution. A naive inversion transform method similar to that in \citet{broadie2006exact} is implemented in \citet{Wu_Simulation_of_Affine_2025}. Given the closeness to normal density shape, fine tunes of the simulation procedure with numerical expertise would probably resulting in a great speed up, however we leave it for these experts.

\subsection{Simulation experiments}
Given the derived conditional moments and unconditional moments of AJDs, the moment-based method introduced above provides an efficient simulation from both the  conditional distribution and the steady-state distribution for these models. Simulation experiments are conducted to validate the efficiency and effectiveness of our proposed method. Conditional distribution simulation experiments are first conducted for the Heston, SVJ and SVCJ models, followed by steady-state distribution experiments for the SVCJ model. The conditional and unconditional moment formulae used in the experiments have already been integrated into our R implementation of the simulation method \citep{Wu_Simulation_of_Affine_2025}, and not presented here due to their lengthy expressions. All simulation experiments are run in R 4.2.2 in macOS 15.3.1 on a machine with a 4.5 GHz Apple M4 Pro processor and 24 GB of RAM.

\subsubsection{Conditional distribution simulation}
We test the accuracy and speed of our simulation method in simulating samples from the conditional distributions of three AJDs. The European call option pricing problem is employed to measure accuracies of the simulation methods. Following the measurement in the literature, the following root-mean-squared (RMS) errors are used: 
$$\mathrm{RMSE} = \sqrt{\frac{1}{G}\sum_{g=1}^G \left(\tilde{P}_g(N) - P\right)^2},$$
where $P$ denotes the true option price, $\tilde{P}_g(N)$ denotes the price estimation from the $g$-th Monte Carlo experiment with $N$ samples, and $G$ denotes the total replications of experiments which is set as 1000 through the experiments. Note that the true option price $P$ is computable from corresponding pricing algorithms in the literature \citep{heston1993closed, bates1996jumps, duffie2000transform}. Striking prices of the options are denoted by $K$ across all the followed tables. All parameter settings within this subsection are adopted from \citet{broadie2006exact}. 

The simulation experiments start under the Heston SV model. Distinct from two existing methods, \citet{broadie2006exact} and \citet{kyriakou2023unified}, our method demands not a single evaluation of the modified Bessel function of the first kind whose evaluation consumes time due to its infinite series expression. The two existing methods demand at least tens of Bessel function evaluations for generating each single sample from the AJD models. Therefore, our method achieves a very fast speed-up for generating samples from the target AJDs, yielding one hundred times more or less faster speed, as shown in Tables~\ref{tab:heston-1} and \ref{tab:heston-2}. In terms of accuracy, the results also show that our method is virtually as accurate as the other two exact simulation schemes.
\begin{table}[!ht]
  \centering
  \begin{threeparttable}
  \caption{Simulation results under the Heston model scenario 1}
  \label{tab:heston-1}
  \setlength{\tabcolsep}{3pt}
  \begin{tabular}{DZZZ|ZZZ|ZZZ}
    \toprule
    &\multicolumn{3}{c|}{Our method} &\multicolumn{3}{c|}{Broadie-Kaya} &\multicolumn{3}{c}{Kyriakou-Brignone-Fusai}\\
    \cline{2-10}
    N ($10^4$) & RMS error & Time &  BesselI  & RMS error & Time\tnote{*} &  BesselI\tnote{*} ($10^4$) & RMS error & Time\tnote{*} & BesselI\tnote{*} ($10^4$) \\
    \midrule
    1 & 0.0733 & 0.09 & - & 0.0750 &41.51 &66 &0.0736  &8.86 &28 \\ 
    4 & 0.0373 & 0.36 & - & 0.0373 &166.04 &4$\times$66 &0.0372  &35.11 &4$\times$28\\
    16 & 0.0182 & 1.41 & - & 0.0186 &664.16 &16$\times$66 &0.0186  &140.45 &16$\times$28\\
    64 & 0.0090 & 5.63 & - & 0.0093 &2656.64 &64$\times$66 &0.0093  &561.79 &64$\times$28\\
    256 & 0.0048 & 22.70 & - & 0.0046 &10626.56 &256$\times$66 &0.0046  &2247.17 &256$\times$28\\
    \bottomrule
  \end{tabular}
  \textit{Note:} Parameters $s_0=100$, $K=100$, $v_0=0.010201$, $k=6.21$,
  $\theta=0.019$, $\sigma_v=0.61$, $\rho=-0.70$, $\mu=3.19$\%, $t=1.0$ (year), true option price $=6.8061$. Time: computing time in seconds. *: The values in these columns for the last four rows are estimated based on the values in the first row.
  \end{threeparttable}
\end{table}
\begin{table}[!ht]
  \centering
  \begin{threeparttable}
  \caption{Simulation results under the Heston Model Scenario 2}
  \label{tab:heston-2}
  \setlength{\tabcolsep}{3pt}
  \begin{tabular}{DZZZ|ZZZ|ZZZ}
    \toprule
    &\multicolumn{3}{c|}{Our method} &\multicolumn{3}{c|}{Broadie-Kaya} &\multicolumn{3}{c}{Kyriakou-Brignone-Fusai}\\
    \cline{2-10}
    N ($10^4$) & RMS error & Time & BesselI  & RMS error & Time\tnote{*} & BesselI\tnote{*} ($10^4$) & RMS error & Time\tnote{*} & BesselI\tnote{*} ($10^4$)\\
    \midrule
    1 & 0.5764 & 0.09 & - & 0.6125 &62.75 &75 &0.5786  &8.90 &28\\ 
    4 & 0.3084 & 0.36 & - & 0.2904 &251.00 &4$\times$75 &0.3000  &35.03 &4$\times$28\\
    16 & 0.1493 & 1.43 & - & 0.1464 &1004.00 &16$\times$75 &0.1440  &140.11 &16$\times$28\\
    64 & 0.0741 & 5.76 & - & 0.0726 &4016.00 &64$\times$75 &0.0734  &560.45 &64$\times$28\\
    256 & 0.0367 & 23.42 & - & 0.0362 &16064.00 &256$\times$75 &0.0367  &2241.79 &256$\times$28\\
    \bottomrule
  \end{tabular}
  \textit{Note:} Parameters $s_0=100$, $K=100$, $v_0=0.09$, $k=2.00$,
  $\theta=0.09$, $\sigma_v=1.00$, $\rho=-0.30$, $\mu=5.00$\%, $t=5.0$ (years), true option price $=34.9998$. Time: computing time in seconds. *: The values in these columns for the last four rows are estimated based on the values in the first row.
  \end{threeparttable}
\end{table}
Similar results are reported in Tables~\ref{tab:svj} and \ref{tab:svcj} for the SVJ and SVCJ models, respectively.
\begin{table}[!ht]
  \centering
  \begin{threeparttable}
  \caption{Simulation results under the SVJ Model}
  \label{tab:svj}
  \setlength{\tabcolsep}{6pt}
  \begin{tabular}{DZZZ|ZZZ}
    \toprule
    &\multicolumn{3}{c|}{Our method} &\multicolumn{3}{c}{Broadie-Kaya}\\
    \cline{2-7}
    N ($10^4$) & RMS error & Time (sec.) & BesselI & RMS error & Time\tnote{*} (sec.) & BesselI\tnote{*} ($10^4$)\\
    \midrule
    1 & 0.2330 & 0.10 & - & 0.2232 &8.34 &35 \\ 
    4 & 0.1083 & 0.38 & - & 0.1124 &33.35 &4$\times$35 \\
    16 & 0.0570 & 1.50 & - & 0.0560 &133.41 &16$\times$35 \\
    64 & 0.0274 & 5.96 & - & 0.0280 &533.63 &64$\times$35 \\
    256 & 0.0135 & 23.83 & - & 0.0140 &2134.53 &256$\times$35 \\
    \bottomrule
  \end{tabular}
  \textit{Note:} Parameters $s_0=100$, $K=100$, $v_0=0.008836$, $k=3.99$,
  $\theta=0.0014$, $\sigma_v=0.27$, $\rho=-0.79$, $\lambda = 0.11$, $\mu_s=-0.139083$, $\sigma_s=0.15$, $\mu=4.51$\%, $t=5.0$ (years), true option price $=20.1642$. *: The values in these columns for the last four rows are estimated based on the values in the first row.
  \end{threeparttable}
\end{table}
\begin{table}[!ht]
  \centering
  \begin{threeparttable}
  \caption{Simulation results under the SVCJ Model}
  \label{tab:svcj}
  \setlength{\tabcolsep}{6pt}
  \begin{tabular}{DZZZ|ZZZ}
    \toprule
    &\multicolumn{3}{c|}{Our method} &\multicolumn{3}{c}{Broadie-Kaya}\\
    \cline{2-7}
    N ($10^4$) & RMS error & Time (sec.) & BesselI  & RMS error & Time\tnote{*} (sec.) & BesselI\tnote{*} ($10^4$)\\
    \midrule
    1 & 0.0716 & 0.09 & - & 0.0720 &11.66 &37 \\ 
    4 & 0.0368 & 0.35 & - & 0.0369 &46.64 &4$\times$37 \\
   16 & 0.0183 & 1.38 & - & 0.0184 &186.56 &16$\times$37 \\
   64 & 0.0087 & 5.52 & - & 0.0092 &746.24 &64$\times$37 \\
  256 & 0.0044 & 22.12 & - & 0.0046 &2984.96 &256$\times$37 \\
    \bottomrule
  \end{tabular}
  \textit{Note:} Parameters $s_0=100$, $K=100$, $v_0=0.007569$, $k=3.46$,
  $\theta=0.008$, $\sigma_v=0.14$, $\rho=-0.82$, $\lambda=0.47$, $\mu_s=-0.086539$, $\sigma_s=0.0001$, $\mu_v=0.05$, $\rho_J=-0.38$, $\mu=7.89$\%, $t=1.0$ (year), true option price $=6.8619$. *: The values in these columns for the last four rows are estimated based on the values in the first row.
  \end{threeparttable}
\end{table}

The existing jump-by-jump updating scheme \citep{broadie2006exact, kyriakou2023unified} for the SVCJ model suffers from a linear growth of computational costs with respect to time horizon $t$ and jump intensity $\lambda$, though Table~\ref{tab:svcj} reports similar results as for the other models under a low-jump setting (mean jump times $=0.47$). Therefore, long time horizon or frequent-jump scenarios would make existing methods highly inefficient. In contrast, our approach remains computationally efficient regardless of time horizon or jump frequency, as evidenced by the stable computation times in Table~\ref{tab:svcj}, which mirror those of the Heston and SVJ cases.

\subsubsection{Steady-state distribution simulation}
A further advantage of our method lies in its ability to generate samples directly from the steady-state distributions of AJDs using the derived unconditional moments. This capability is particularly valuable for AJDs having jumps in the latent process, such as the SVCJ model, where the steady-state distribution of the latent process lacks an analytical form.

For AJDs with a square-root diffusion latent process, steady-state simulation can be achieved via a two-step procedure: (1) sampling the latent process from its gamma steady-state distribution, followed by (2) conditional simulation using one of the three established methods. However, this approach incurs more computational overhead and cannot be directly extended to AJDs having jumps in the latent process. 
Our method overcomes these limitations by providing a unified framework for steady-state simulation. Through moment-based density approximation, we enable direct and efficient sampling even for jump-extended latent processes, offering both theoretical generality and computational advantages.

We evaluate our method's performance through steady-state distribution simulations under the SVCJ model, using the same parameters as in Table~\ref{tab:svcj} but with modified option valuation targets. To assess simulation accuracy, we compare against an unconditional option price benchmark $e^{-rt}\mathbb{E}[(s_0 e^{y_t} - K)^{+}]$ which differs from the standard conditional expectation $e^{-rt}\E[(s_0 e^{y_t} - K)^{+}|v_0]$, where $r$ denotes the risk-less rate. The true unconditional value is computed via a two-step method: 
\begin{enumerate}
    \item Variance stabilization: Simulate the variance process for an extended period from any positive initial value until reaching steady state.
    \item Option valuation: Compute the option price using the stabilized variance sample.
\end{enumerate}
This procedure is repeated extensively to generate a high-precision benchmark as the sample mean. As demonstrated in Table~\ref{tab:svcj-steady-state} (combined with Table~\ref{tab:svcj} results), our method achieves comparable accuracy and computational efficiency for both steady-state and conditional distribution simulations.

\begin{table}[!ht]
  \centering
  \begin{threeparttable}
  \caption{Steady-state distribution simulation for the SVCJ Model}
  \label{tab:svcj-steady-state}
  \setlength{\tabcolsep}{8pt}
  \begin{tabular}{DZZ}
    \toprule
    N ($10^4$) & RMS error & Time (seconds) \\
    \midrule
    1 & 0.0795 & 0.09  \\ 
    4 & 0.0395 & 0.34  \\
   16 & 0.0197 & 1.37  \\
   64 & 0.0099 & 5.48  \\
  256 & 0.0051 & 21.92 \\
    \bottomrule
  \end{tabular}
  \textit{Note:} The parameters are consistent with those in Table~\ref{tab:svcj}, with the true value set as the ``unconditional" option price $e^{-rt}\E[(s_0e^{y_t} - K)^{+}] =7.109439$ ($r=3.19\%$).
  \end{threeparttable}
\end{table}

Before closing the section, we emphasize that when European option pricing is the primary objective, our method offers a computatiobally efficient alternative to simulation-based approaches. The accurately approximated density obtained through our framework enable direct computation of option prices via numerical integration of $e^{-rt}\E[(s_0e^{y_t}-K)^{+}|v_0]$. Consequently, our approach serves as a competitive alternative to established Fourier-based pricing methods, including those developed by \citet{heston1993closed} for the Heston model, \citet{bates1996jumps} for SVJ specifications, and \citet{duffie2000transform} for SVCJ frameworks.

\section{Conclusion}
We develop a recursive methodology for deriving closed-form solutions to both conditional and unconditional moments of affine jump diffusions with state-independent jump intensities. The derived moments enable the construction of closed-form density approximations (up to a normalization constant) for both conditional and steady-state distributions of such processes. The availability of these closed-form moments and analytical density approximations significantly enhances the tractability of the target AJDs, facilitating important applications, such as European option pricing via direct numerical integration, fast exact simulation and straightforward parameter estimation via method of moments.

Furthermore, our recursive framework offers additional promising extensions. The joint moments of intermediate variables could potentially enable joint distribution approximations, extending our approach to sample path simulation. Future research directions may also include generalization to state-dependent jump intensities and adaptation to non-affine processes.

\section*{Acknowledgement}
This work is supported in part by the National Nature Science Foundation of China (NSFC) under grants 72033003, 72350710219 and 72342006.

\bibliography{references}

\begin{thebibliography}{}

\bibitem[A{\"\i}t-Sahalia and Kimmel, 2007]{ai2007maximum}
A{\"\i}t-Sahalia, Y. and Kimmel, R. (2007).
\newblock Maximum likelihood estimation of stochastic volatility models.
\newblock {\em Journal of financial economics}, 83(2):413--452.

\bibitem[Barczy et~al., 2018]{barczy2018asymptotic}
Barczy, M., Alaya, M.~B., Kebaier, A., and Pap, G. (2018).
\newblock Asymptotic properties of maximum likelihood estimator for the growth
  rate for a jump-type cir process based on continuous time observations.
\newblock {\em Stochastic Processes and their Applications}, 128(4):1135--1164.

\bibitem[Bates, 1996]{bates1996jumps}
Bates, D.~S. (1996).
\newblock Jumps and stochastic volatility: Exchange rate processes implicit in
  deutsche mark options.
\newblock {\em The Review of Financial Studies}, 9(1):69--107.

\bibitem[Bollerslev and Zhou, 2002]{bollerslev2002estimating}
Bollerslev, T. and Zhou, H. (2002).
\newblock Estimating stochastic volatility diffusion using conditional moments
  of integrated volatility.
\newblock {\em Journal of Econometrics}, 109(1):33--65.

\bibitem[Broadie and Kaya, 2006]{broadie2006exact}
Broadie, M. and Kaya, {\"O}. (2006).
\newblock Exact simulation of stochastic volatility and other affine jump
  diffusion processes.
\newblock {\em Operations Research}, 54(2):217--231.

\bibitem[Cai et~al., 2017]{cai2017exact}
Cai, N., Song, Y., and Chen, N. (2017).
\newblock Exact simulation of the sabr model.
\newblock {\em Operations Research}, 65(4):931--951.

\bibitem[Choudhury and Lucantoni, 1996]{choudhury1996numerical}
Choudhury, G.~L. and Lucantoni, D.~M. (1996).
\newblock Numerical computation of the moments of a probability distribution
  from its transform.
\newblock {\em Operations Research}, 44(2):368--381.

\bibitem[Cox et~al., 1985]{cox1985theory}
Cox, J.~C., Ingersoll~Jr, J.~E., and Ross, S.~A. (1985).
\newblock A theory of the term structure of interest rates.
\newblock {\em Econometrica}, 53:385--407.

\bibitem[Cuchiero et~al., 2012]{cuchiero2012polynomial}
Cuchiero, C., Keller-Ressel, M., and Teichmann, J. (2012).
\newblock Polynomial processes and their applications to mathematical finance.
\newblock {\em Finance and Stochastics}, 16:711--740.

\bibitem[Cui et~al., 2021]{cui2021efficient}
Cui, Z., Kirkby, J.~L., and Nguyen, D. (2021).
\newblock Efficient simulation of generalized sabr and stochastic local
  volatility models based on markov chain approximations.
\newblock {\em European Journal of Operational Research}, 290(3):1046--1062.

\bibitem[Dassios and Zhao, 2017]{dassios2017efficient}
Dassios, A. and Zhao, H. (2017).
\newblock Efficient simulation of clustering jumps with cir intensity.
\newblock {\em Operations Research}, 65(6):1494--1515.

\bibitem[Duffie et~al., 2003]{duffie2003affine}
Duffie, D., Filipovi{\'c}, D., and Schachermayer, W. (2003).
\newblock Affine processes and applications in finance.
\newblock {\em The Annals of Applied Probability}, 13(3):984--1053.

\bibitem[Duffie et~al., 2000]{duffie2000transform}
Duffie, D., Pan, J., and Singleton, K. (2000).
\newblock Transform analysis and asset pricing for affine jump-diffusions.
\newblock {\em Econometrica}, 68(6):1343--1376.

\bibitem[Filipovi{\'c} and Larsson, 2016]{filipovic2016polynomial}
Filipovi{\'c}, D. and Larsson, M. (2016).
\newblock Polynomial diffusions and applications in finance.
\newblock {\em Finance and Stochastics}, 20(4):931--972.

\bibitem[Filipovi{\'c} and Larsson, 2020]{filipovic2020polynomial}
Filipovi{\'c}, D. and Larsson, M. (2020).
\newblock Polynomial jump-diffusion models.
\newblock {\em Stochastic Systems}, 10(1):71--97.

\bibitem[Filipovic and Mayerhofer, 2009]{filipovic2009affine}
Filipovic, D. and Mayerhofer, E. (2009).
\newblock Affine diffusion processes: theory and applications.
\newblock {\em Advanced financial modelling}, 8:1--40.

\bibitem[Giesecke et~al., 2011]{giesecke2011exact}
Giesecke, K., Kakavand, H., and Mousavi, M. (2011).
\newblock Exact simulation of point processes with stochastic intensities.
\newblock {\em Operations research}, 59(5):1233--1245.

\bibitem[Glasserman and Kim, 2010]{glasserman2010moment}
Glasserman, P. and Kim, K.-K. (2010).
\newblock Moment explosions and stationary distributions in affine diffusion
  models.
\newblock {\em Mathematical Finance}, 20(1):1--33.

\bibitem[Glasserman and Kim, 2011]{glasserman2011gamma}
Glasserman, P. and Kim, K.-K. (2011).
\newblock Gamma expansion of the heston stochastic volatility model.
\newblock {\em Finance and Stochastics}, 15:267--296.

\bibitem[Heston, 1993]{heston1993closed}
Heston, S.~L. (1993).
\newblock A closed-form solution for options with stochastic volatility with
  applications to bond and currency options.
\newblock {\em The Review of Financial Studies}, 6(2):327--343.

\bibitem[Jiang and Knight, 2002]{jiang2002estimation}
Jiang, G.~J. and Knight, J.~L. (2002).
\newblock Estimation of continuous-time processes via the empirical
  characteristic function.
\newblock {\em Journal of Business \& Economic Statistics}, 20(2):198--212.

\bibitem[Jin et~al., 2016]{jin2016positive}
Jin, P., R{\"u}diger, B., and Trabelsi, C. (2016).
\newblock Positive harris recurrence and exponential ergodicity of the basic
  affine jump-diffusion.
\newblock {\em Stochastic Analysis and Applications}, 34(1):75--95.

\bibitem[Kallsen and Muhle-Karbe, 2010]{kallsen2010exponentially}
Kallsen, J. and Muhle-Karbe, J. (2010).
\newblock Exponentially affine martingales, affine measure changes and
  exponential moments of affine processes.
\newblock {\em Stochastic Processes and their Applications}, 120(2):163--181.

\bibitem[Kang et~al., 2017]{kang2017exact}
Kang, C., Kang, W., and Lee, J.~M. (2017).
\newblock Exact simulation of the wishart multidimensional stochastic
  volatility model.
\newblock {\em Operations Research}, 65(5):1190--1206.

\bibitem[Keller-Ressel and Mayerhofer, 2015]{keller2015}
Keller-Ressel, M. and Mayerhofer, E. (2015).
\newblock Exponential moments of affine processes.
\newblock {\em The annals of applied probability}, 25(2):714--752.

\bibitem[Kyriakou et~al., 2024]{kyriakou2023unified}
Kyriakou, I., Brignone, R., and Fusai, G. (2024).
\newblock Unified moment-based modeling of integrated stochastic processes.
\newblock {\em Operations Research}, 72(4):1630--1653.

\bibitem[Li and Wu, 2019]{li2019exact}
Li, C. and Wu, L. (2019).
\newblock Exact simulation of the ornstein--uhlenbeck driven stochastic
  volatility model.
\newblock {\em European Journal of Operational Research}, 275(2):768--779.

\bibitem[Overbeck and Rydén, 1997]{overbeck1997}
Overbeck, L. and Rydén, T. (1997).
\newblock Estimation in the cox-ingersoll-ross model.
\newblock {\em Econometric Theory}, 13(3):430--461.

\bibitem[Rose and Smith, 2002]{rose2002mathstatica}
Rose, C. and Smith, M.~D. (2002).
\newblock {\em Mathstatica: mathematical statistics with mathematica}.
\newblock Springer-Verlag.

\bibitem[Shreve, 2004]{shreve2004stochastic}
Shreve, S.~E. (2004).
\newblock {\em Stochastic Calculus for Finance II: Continuous-Time Models},
  volume~11.
\newblock Springer Science \& Business Media.

\bibitem[Wu, 2025]{Wu_Simulation_of_Affine_2025}
Wu, Y.-F. (2025).
\newblock ajd.sim.wh: An r package for simulating affine jump-diffusions via
  the wu-hu method.
\newblock https://github.com/xmlongan/ajd.sim.wh.

\bibitem[Wu and Hu, 2025]{wu2024ajdmom}
Wu, Y.-F. and Hu, J.-Q. (2025).
\newblock ajdmom: A python package for deriving moment formulas of affine jump
  diffusion processes.

\bibitem[Wu et~al., 2024]{wu2023method}
Wu, Y.-F., Yang, X., and Hu, J.-Q. (2024).
\newblock Method of moments estimation for affine stochastic volatility models.

\end{thebibliography}

\appendix

\section*{Appendix A: Moment examples for the Heston model}
We provide some low-order moment formula examples for the Heston model to demonstrate their structures. Moments for all other AJDs under consideration take similar expressions, and can be easily derived by the Python package \emph{ajdmom} which is readily available for installation from the Python package index (PyPI).

The (unconditional) central moment formulae presented in Tables~\ref{tab:hest-cmom2} - \ref{tab:hest-cmom4} are decoded as:
\begin{equation}\label{eqn:central-moment-heston}
    \mathbb{E}[\bar{y}_t^m] = \sum_{\boldsymbol{i}}e^{-i_1kt}t^{i_2}(1/k)^{i_3}\theta^{i_4}\sigma_v^{i_5}\rho^{i_6} \times \frac{i_7}{i_8},
\end{equation}
where $\boldsymbol{i}=(i_1,\dots,i_8)$ represents each row of numbers in the tables. Please kindly note that each row in these tables actually contains two rows of numbers.

\renewcommand{\arraystretch}{0.6}
\setlength{\tabcolsep}{6pt}
\begin{table}[ht]
\caption{The second central moment of the Heston model, $\mathbb{E}[\bar{y}_t^2]$.}
\label{tab:hest-cmom2}
\centering
\begin{tabular}{ZZZZ ZZDD|ZZZZ ZZDD}
\toprule
$i_{1}$ & $i_{2}$ & $i_{3}$ & $i_{4}$ & $i_{5}$ & $i_{6}$ & $i_{7}$ & $i_{8}$ &
$i_{1}$ & $i_{2}$ & $i_{3}$ & $i_{4}$ & $i_{5}$ & $i_{6}$ & $i_{7}$ & $i_{8}$\\
\midrule
0 & 0 & 3 & 1 & 2 & 0 & -1 & 4 & 1 & 0 & 3 & 1 & 2 & 0 & 1 & 4\\
0 & 0 & 2 & 1 & 1 & 1 & 1 & 1 & 1 & 0 & 2 & 1 & 1 & 1 & -1 & 1\\
0 & 1 & 1 & 1 & 1 & 1 & -1 & 1 & 0 & 1 & 2 & 1 & 2 & 0 & 1 & 4\\
0 & 1 & 0 & 1 & 0 & 0 & 1 & 1 &  &  &  &  &  &  &  &\\ 
\bottomrule
\end{tabular}
\end{table}

\begin{table}[ht]
\caption{The third central moment of the Heston model, $\mathbb{E}[\bar{y}_t^3]$.}
\label{tab:hest-cmom3}
\centering
\begin{tabular}{ZZZZ ZZDD|ZZZZ ZZDD}
\toprule
$i_{1}$ & $i_{2}$ & $i_{3}$ & $i_{4}$ & $i_{5}$ & $i_{6}$ & $i_{7}$ & $i_{8}$ &
$i_{1}$ & $i_{2}$ & $i_{3}$ & $i_{4}$ & $i_{5}$ & $i_{6}$ & $i_{7}$ & $i_{8}$\\
\midrule
0 & 0 & 5 & 1 & 4 & 0 & 3 & 4 & 1 & 0 & 5 & 1 & 4 & 0 & -3 & 4\\
1 & 1 & 3 & 1 & 3 & 1 & 9 & 4 & 1 & 1 & 4 & 1 & 4 & 0 & -3 & 8\\
1 & 0 & 3 & 1 & 2 & 2 & -6 & 1 & 1 & 0 & 4 & 1 & 3 & 1 & 9 & 2\\
0 & 0 & 3 & 1 & 2 & 2 & 6 & 1 & 0 & 0 & 4 & 1 & 3 & 1 & -9 & 2\\
1 & 0 & 3 & 1 & 2 & 0 & -3 & 2 & 0 & 0 & 3 & 1 & 2 & 0 & 3 & 2\\
1 & 1 & 2 & 1 & 2 & 2 & -3 & 1 & 0 & 1 & 2 & 1 & 2 & 2 & -3 & 1\\
0 & 1 & 3 & 1 & 3 & 1 & 9 & 4 & 0 & 1 & 4 & 1 & 4 & 0 & -3 & 8\\
0 & 1 & 1 & 1 & 1 & 1 & 3 & 1 & 0 & 1 & 2 & 1 & 2 & 0 & -3 & 2\\
1 & 0 & 2 & 1 & 1 & 1 & 3 & 1 & 0 & 0 & 2 & 1 & 1 & 1 & -3 & 1\\
\bottomrule
\end{tabular}
\end{table}

\begin{longtable}{ZZZZ ZZDD|ZZZZ ZZDD}
\caption{The fourth central moment of the Heston model, $\mathbb{E}[\bar{y}_t^4]$.}
\label{tab:hest-cmom4}\\[-12pt]
\toprule
$i_{1}$ & $i_{2}$ & $i_{3}$ & $i_{4}$ & $i_{5}$ & $i_{6}$ & $i_{7}$ & $i_{8}$ &
$i_{1}$ & $i_{2}$ & $i_{3}$ & $i_{4}$ & $i_{5}$ & $i_{6}$ & $i_{7}$ & $i_{8}$\\
\midrule
\endfirsthead
\toprule
$i_{1}$ & $i_{2}$ & $i_{3}$ & $i_{4}$ & $i_{5}$ & $i_{6}$ & $i_{7}$ & $i_{8}$ &
$i_{1}$ & $i_{2}$ & $i_{3}$ & $i_{4}$ & $i_{5}$ & $i_{6}$ & $i_{7}$ & $i_{8}$\\
\midrule
\endhead
\midrule
\multicolumn{16}{c}{Continued on the next page}\\
\bottomrule
\endfoot
\bottomrule
\endlastfoot
0 & 0 & 6 & 2 & 4 & 0 & 3 & 16 & 1 & 0 & 6 & 2 & 4 & 0 & -3 & 8\\
2 & 0 & 6 & 2 & 4 & 0 & 3 & 16 & 1 & 1 & 4 & 2 & 3 & 1 & -3 & 1\\
1 & 1 & 5 & 2 & 4 & 0 & 3 & 8 & 1 & 0 & 4 & 2 & 2 & 2 & -6 & 1\\
1 & 0 & 5 & 2 & 3 & 1 & 3 & 1 & 2 & 0 & 4 & 2 & 2 & 2 & 3 & 1\\
2 & 0 & 5 & 2 & 3 & 1 & -3 & 2 & 0 & 0 & 4 & 2 & 2 & 2 & 3 & 1\\
0 & 0 & 5 & 2 & 3 & 1 & -3 & 2 & 1 & 1 & 3 & 2 & 2 & 2 & 6 & 1\\
1 & 1 & 3 & 2 & 2 & 0 & 3 & 2 & 0 & 1 & 3 & 2 & 2 & 2 & -6 & 1\\
0 & 1 & 4 & 2 & 3 & 1 & 3 & 1 & 0 & 1 & 5 & 2 & 4 & 0 & -3 & 8\\
0 & 1 & 2 & 2 & 1 & 1 & 6 & 1 & 0 & 1 & 3 & 2 & 2 & 0 & -3 & 2\\
1 & 1 & 2 & 2 & 1 & 1 & -6 & 1 & 0 & 0 & 7 & 1 & 6 & 0 & -87 & 32\\
1 & 0 & 7 & 1 & 6 & 0 & 21 & 8 & 2 & 0 & 7 & 1 & 6 & 0 & 3 & 32\\
1 & 1 & 5 & 1 & 5 & 1 & -15 & 1 & 1 & 1 & 6 & 1 & 6 & 0 & 15 & 8\\
1 & 0 & 5 & 1 & 4 & 2 & 51 & 1 & 1 & 0 & 6 & 1 & 5 & 1 & -21 & 1\\
2 & 0 & 5 & 1 & 4 & 2 & 3 & 2 & 2 & 0 & 6 & 1 & 5 & 1 & -3 & 4\\
0 & 0 & 5 & 1 & 4 & 2 & -105 & 2 & 0 & 0 & 6 & 1 & 5 & 1 & 87 & 4\\
1 & 0 & 5 & 1 & 4 & 0 & 9 & 1 & 0 & 0 & 5 & 1 & 4 & 0 & -9 & 1\\
1 & 1 & 4 & 1 & 4 & 2 & 36 & 1 & 1 & 2 & 3 & 1 & 4 & 2 & 15 & 2\\
1 & 2 & 4 & 1 & 5 & 1 & -3 & 1 & 1 & 2 & 5 & 1 & 6 & 0 & 3 & 8\\
1 & 1 & 4 & 1 & 4 & 0 & 9 & 2 & 1 & 1 & 3 & 1 & 3 & 3 & -24 & 1\\
1 & 0 & 4 & 1 & 3 & 3 & -36 & 1 & 0 & 0 & 4 & 1 & 3 & 3 & 36 & 1\\
1 & 1 & 3 & 1 & 3 & 1 & -18 & 1 & 1 & 0 & 4 & 1 & 3 & 1 & -36 & 1\\
0 & 0 & 4 & 1 & 3 & 1 & 36 & 1 & 1 & 2 & 2 & 1 & 3 & 3 & -6 & 1\\
0 & 2 & 2 & 2 & 2 & 2 & 3 & 1 & 0 & 2 & 3 & 2 & 3 & 1 & -3 & 2\\
0 & 2 & 4 & 2 & 4 & 0 & 3 & 16 & 0 & 1 & 3 & 1 & 3 & 3 & -12 & 1\\
0 & 1 & 4 & 1 & 4 & 2 & 18 & 1 & 0 & 1 & 5 & 1 & 5 & 1 & -15 & 2\\
0 & 1 & 6 & 1 & 6 & 0 & 15 & 16 & 0 & 0 & 3 & 1 & 2 & 2 & -24 & 1\\
1 & 0 & 3 & 1 & 2 & 2 & 24 & 1 & 0 & 2 & 1 & 2 & 1 & 1 & -6 & 1\\
0 & 2 & 2 & 2 & 2 & 0 & 3 & 2 & 0 & 1 & 2 & 1 & 2 & 2 & 12 & 1\\
0 & 1 & 3 & 1 & 3 & 1 & -18 & 1 & 0 & 1 & 4 & 1 & 4 & 0 & 9 & 2\\
1 & 1 & 2 & 1 & 2 & 2 & 12 & 1 & 0 & 0 & 3 & 1 & 2 & 0 & -3 & 1\\
1 & 0 & 3 & 1 & 2 & 0 & 3 & 1 & 0 & 2 & 0 & 2 & 0 & 0 & 3 & 1\\
0 & 1 & 2 & 1 & 2 & 0 & 3 & 1 &  &  &  &  &  &  &  & \\
\end{longtable}

\section*{Appendix B: Joint moments of contemporaneous CPPs}\label{app:joint-moments-cpps}
In this section, we present the computation of the joint moments of ($\iez_{s,t}, \iz_{s,t}, \iz_{s,t}^*$). This is accomplished by deriving the joint moment-generating function (MGF) of these contemporaneous CPP processes.

The joint MGF of ($\iez_{s,t}, \iz_{s,t}, \iz_{s,t}^*$) can be derived as follows:
\begin{align*}
    M_{\iez_{s,t},\iz_{s,t},\iz_{s,t}^*}(\mathbf{a})
    &\defeq \E[e^{a_1\iez_{s,t} + a_2\iz_{s,t} + a_3\iz_{s,t}^*}]\\
    &= \E[\E[e^{\sum_{i=1}^n [(a_1 e^{ks_i} + a_2)J_i + a_3J_i^*]} |N(t-s) = n]]\\
    &= \E[\E[\left(\E[M_{J_i}(a_1e^{ks_i} + a_2)]\cdot M_{J_i^*}(a_3)\right)^n |N(t-s) = n]]\\
    &= \sum_{n=0}^{\infty} \left(\E[M_{J_i}(a_1e^{ks_i} + a_2)] \cdot M_{J_i^*}(a_3)\right)^n \frac{[\lambda (t-s)]^n e^{-\lambda (t-s)}}{n!}\\
    &= e^{-\lambda (t-s)}  \sum_{n=0}^{\infty}  \left[\lambda (t-s)\E[M_{J_i}(a_1e^{ks_i} + a_2)]\cdot M_{J_i^*}(a_3)\right]^n/n!\\
    &= e^{\lambda (t-s)\left(\E[M_{J_i}(a_1e^{ks_i} + a_2)]\cdot M_{J_i^*}(a_3) - 1\right)},
\end{align*}
where $\mathbf{a}$ is a vector of real numbers, i.e., $\mathbf{a} \defeq (a_1, a_2, a_3)$, $N(\cdot)$ is the shared counting process for ($\iez_{s,t}, \iz_{s,t}, \iz_{s,t}^*$), and $M_{J_i}(\cdot)$ and $M_{J_i^*}(\cdot)$ are the MGFs of $J_i^*$ and $J_i$, respectively. Note that, conditional on $N(t-s) = n$, the unsorted arrival times are uniformly distributed over the interval $(s,t]$. For notational simplicity, we use $\{s_1,\dots, s_n\}$ to denote these unsorted arrival times and omit the conditional notation $|N(t-s) = n$ in the expectation.  In what follows, we will demonstrate that $\E[M_{J_i}(a_1e^{ks_i} + a_2)]$ admits a closed-form expression.

First, we substitute the MGF of $J_i$ (conditioned on $s_i$) with its known formula:
\begin{align*}
    \E[M_{J_i}(a_1e^{ks_i} + a_2)]
    = \E[\E[e^{(a_1e^{ks_i} + a_2)J_i}|s_i]]
    = \E\left[\frac{1}{1-(a_1e^{ks_i} + a_2 )\mu_v} \right].
\end{align*}
By introducing two new variables $a \defeq - a_1\mu_v, b \defeq 1 - a_2\mu_v$, the above expectation simplifies to:
\begin{align*}
    \E[M_{J_i}(a_1e^{ks_i} + a_2)]
    = \E\left[\frac{1}{ae^{ks_i} + b}\right]
    = \frac{1}{t-s}\int_{s}^t\frac{1}{ae^{ks_i}+b}\d s_i.
\end{align*}
The integral can be computed explicitly using the variable substitution method. Let us introduce $x \defeq e^{ks_i}$. Then, $s_i = (1/k)\log x$, $\d s_i = [1/(kx)]\d x$. The integral thus becomes:
\begin{align*}
    \int_{s}^t\frac{1}{ae^{ks_i}+b}\d s_i
    = \int_{e^{ks}}^{e^{kt}} \frac{1}{ax + b} \frac{1}{k} \frac{1}{x} \d x
    = \frac{1}{k}\int_{e^{ks}}^{e^{kt}} \frac{1}{ax + b} \frac{1}{x} \d x.
\end{align*}
If $a = 0$, i.e., $a_1 = 0$, the integral simplifies to:
\begin{equation*}
    \int_{s}^t\frac{1}{ae^{ks_i}+b}\d s_i = \frac{1}{b}(t-s).
\end{equation*}
Otherwise, for $a_1\neq 0$, in any neighborhood of the origin of the vector $(a_1,a_2)$ such that  $ax + b \approx 1$, we have:
\begin{align*}
    \int_{e^{ks}}^{e^{kt}} \frac{1}{ax + b} \frac{1}{x} \d x
    &= -\frac{1}{b}\left[\log(ae^{kt} + b) - \log(ae^{ks} + b)\right] + \frac{1}{b}k(t-s).
\end{align*}
Therefore, the integral evaluates to: 
\begin{align*}
    \int_{s}^t\frac{1}{ae^{ks_i}+b}\d s_i
    &= -\frac{1}{kb}\left[\log(ae^{kt} + b) - \log(ae^{ks} + b)\right] + \frac{1}{b}(t-s),
\end{align*}
provided that $(a_1,a_2)$ lies within a sufficiently small neighborhood of the origin $(0,0)$. Finally, we obtain the closed-form expression for $\E[M_{J_i}(a_1e^{ks_i} + a_2)]$ as:
\begin{equation}\label{eqn:mgf-J}
    \E[M_{J_i}(a_1e^{ks_i} + a_2)]
    = \frac{1}{b}\left[1 -\frac{1}{k(t-s)}\left(\log(ae^{kt} + b) - \log(ae^{ks} + b)\right)\right],
\end{equation}
where $a = -a_1\mu_v$, $b = 1 - a_2\mu_v$, and $(a_1,a_2)$ is restricted to a small neighborhood around the origin $(0,0)$. 

To simplify the notation, we define a new function $M_{E\!J,J,J^*}(\cdot)$ as the following product:
\begin{equation*}
    M_{E\!J,J,J^*}(\mathbf{a}) \defeq \E[M_{J_i}(a_1e^{ks_i} + a_2)]\cdot M_{J_i^*}(a_3).
\end{equation*}
We know that the MGF of $J_i^*$ has the following expression:
\begin{equation*}
    M_{J_i^*}(a_3) = e^{\mu_s a_3 + \sigma_s^2a_3^2/2}
\end{equation*}
since $J_I^*$ is normally distributed with mean $\mu_s$ and variance $\sigma_s^2$. Combining these results, we obtain the following closed-form expression for the joint MGF of $(\iez_{s,t}, \iz_{s,t}, \iz_{s,t}^*)$:
\begin{equation}\label{eqn:mgf-iez-iz-iz}
    M_{\iez_{s,t},\iz_{s,t},\iz_{s,t}^*}(\mathbf{a})
    = e^{\lambda (t-s) (M_{E\!J,J,J^*}(\mathbf{a})-1)},
\end{equation}
where $\mathbf{a} = (a_1,a_2,a_3)$, $a = -a_1\mu_v$, $b = 1-a_2\mu_v$, and
\begin{equation*}
    M_{E\!J,J,J^*}(\mathbf{a})
    = \frac{1}{b} \left[ 1 - \frac{1}{k(t-s)}\left(\log(ae^{kt} + b) - \log(ae^{ks} + b)\right) \right] e^{\mu_s a_3 + \sigma_s^2 a_3^2/2}.
\end{equation*}
Given Equation~\eqref{eqn:mgf-iez-iz-iz}, we can compute the joint moment of $(\iez_{s,t}, \iz_{s,t},\iz_{s,t}^*)$ of any order.

The $n$-th ($n\ge1$) partial derivative of $M_{E\!J,J,J^*}(\mathbf{a})$ with respect to $a_1$ is given by
\begin{equation*}
    \frac{\partial^nM_{E\!J,J,J^*}}{\partial a_1^{n}}
    =  \frac{1}{b}\left[\frac{e^{nkt}}{(ae^{kt}+b)^{n}}  - \frac{e^{nks}}{(ae^{ks}+b)^{n}} \right] \frac{(n-1)! \mu_v^n}{k(t-s)} M_{J_i^*}(a_3).
\end{equation*}
Consequently, the $n$-th moment of $E\!J$ can be expressed as
\begin{equation*}
    \E[(e^{ks_i}J_i)^n] = (e^{nkt} - e^{nks})\frac{1}{k(t-s)}(n-1)!\mu_v^n.
\end{equation*}

For $n_1\ge 1$ and $n_2\ge 1$, the following formula holds:
\begin{equation*}
    \frac{\partial^{n_1+n_2}M_{E\!J,J,J^*}}{\partial a_1^{n_1}\partial a_2^{n_2}}
    = \sum_{i=0}^{n_2} \frac{c(n_1,n_2,i)}{b^{n_2-i+1}}\left[\frac{e^{n_1kt}}{(ae^{kt}+b)^{n_1+i}} - \frac{e^{n_1ks}}{(ae^{ks}+b)^{n_1+i}} \right] \frac{\mu_v^{n_1+n_2}}{k(t-s)} M_{J_i^*}(a_3), 
\end{equation*}
where $c(n_1,n_2,i) \defeq n_2!(n_1 - 1 + i)!/(i!)$. An alternative approach involves directly computing the joint moment directly, yielding:
\begin{equation*}
    \E[(e^{ks_i}J_i)^{n_1}J_i^{n_2}] = \E[e^{n_1ks_i}]\E[J_i^{n_1+n_2}] = \frac{1}{n_1k(t-s)}(e^{n_1kt} - e^{n_1ks})(n_1+n_2)!\mu_v^{n_1+n_2}.
\end{equation*}

For $n\ge 1$, we have:
\begin{align*}
    \frac{\partial^{n}M_{E\!J,J,J^*}}{\partial a_2^{n}}
    &= \frac{n!}{b^{n+1}}  \left[ 1 - \frac{1}{k(t-s)}\left(\log(ae^{kt} + b) - \log(ae^{ks} + b)\right) \right]\mu_v^n M_{J_i^*}(a_3)\\
    &\quad + \sum_{i=1}^n\binom{n}{i}\frac{(n-i)!(i-1)!}{b^{n-i+1}} \left[\frac{1}{(ae^{kt}+b)^i} - \frac{1}{(ae^{ks} + b)^i}\right]\frac{1}{k(t-s)} \mu_v^n M_{J_I^*}(a_3)\\
    &= \frac{n!}{b^{n+1}}  \left[ 1 - \frac{1}{k(t-s)}\left(\log(ae^{kt} + b) - \log(ae^{ks} + b)\right) \right]\mu_v^n M_{J_i^*}(a_3)\\
    &\quad + \sum_{i=1}^n\frac{n!/i}{b^{n-i+1}} \left[\frac{1}{(ae^{kt}+b)^i} - \frac{1}{(ae^{ks} + b)^i}\right]\frac{1}{k(t-s)} \mu_v^n M_{J_I^*}(a_3).
\end{align*}
Thus, the $n$-th moment of $J$ is given by $\E[J_i^n] = n!\mu_v^n$.

\section*{Appendix C: Unconditional moments of the SVCJ model}
\label{app:unconditional-moment-v_t}
We have demonstrated that the conditional moment of the return, $\E[y_t^m|v_0]$, is a polynomial in $v_0$. To compute the unconditional moments of $y_t$, we first need to investigate how to calculate the unconditional moments of the variance $v_t$. We assume that the initial variance $v_0$ follows the stationary distribution of $v_t$, and the process $v_t$ is strictly stationary. This implies that $v_0\overset{d}{=} v(t)$, where $\overset{d}{=}$ denotes equality in distribution \citep{glasserman2010moment, jin2016positive}. We use $v$ to denote a random variable from this stationary distribution.

The solution (Equation~\eqref{eqn:svcj-variance-solution1}) to the variance process can be rewritten as:
\begin{equation}\label{eqn:v_t-srdj}
    e^{kt}(v_t-\theta) = (v_0-\theta) + \sigma_v \ie_t + \iez_t.
\end{equation}
The first unconditional moment is calculated as:  
$$\E[v] = \theta + \lambda\mu_v/k,$$ 
since $\E[\iez_t] = \lambda\mu_v (e^{kt}-1)/k$, $\E[\ie_t] = 0$ and $\E[v_t] = \E[v_0] = \E[v]$. This result allows us to rewrite Equation~\eqref{eqn:v_t-srdj} in the following form:
\begin{equation*}
    e^{kt}(v_t-\E[v]) = \sigma_v\ie_t + \overline{\iez}_t + (v_0-\E[v]),
\end{equation*}
where $\overline{\iez}_t \defeq \iez_t - \E[\iez_t]$ represents the centralized term. This centralized term can be decomposed similarly: $\overline{\iez}_t = \overline{\iez}_{s} + \overline{\iez}_{s,t}$ where $\overline{\iez}_{s,t} \defeq \iez_{s,t} - \E[\iez_{s,t}]$. It is straightforward to verify that $\E[\overline{\iez}_{s,t}^m]$ can be expressed as a ``polynomial'':
\begin{equation*}
    \E[\overline{\iez}_{s,t}^m] = \sum_{\mathbf{j}} c_{\mathbf{j}} e^{j_1kt}e^{j_2ks}k^{-j_3}\lambda^{j_4}\mu_v^{j_5},
\end{equation*}
where, with a slight abuse of notation, $\mathbf{j}\defeq (j_1,\dots,j_5)$, and $c_{\mathbf{j}}$ denotes the associated coefficient for the corresponding monomial. Therefore, the conditional joint moment $\E[\ie_t^{m_1}\overline{\iez}_t^{m_2}|v_0]$ can be computed using the following recursive equation:
\begin{equation*}
    \E[\ie_t^{m_1}\overline{\iez}_t^{m_2}|v_0]
    = \sum_{i=0}^{m_2}\binom{m_2}{i}\sum_{\mathbf{j}}c_{\mathbf{j}} e^{j_1kt}k^{-j_3}\lambda^{j_4}\mu_v^{j_5}P(m_1,m_2),\quad m_1 \ge 2,
\end{equation*}
where $P(m_1,m_2) \defeq [m_1(m_1-1)/2]\cdot(p_1 + p_2 + p_3 + p_4)$, and 
\begin{align*}
    p_1 &\defeq \int_0^t e^{(j_2+1)ks}\E[\ie_s^{m_1-2}\overline{\iez}_s^i|v_0] \d s \times (v_0-\E[v]),\\
    p_2 &\defeq \int_0^t e^{(j_2+2)ks}\E[\ie_s^{m_1-2}\overline{\iez}_s^i|v_0] \d s \times \E[v],\\
    p_3 &\defeq \int_0^t e^{(j_2+1)ks}\E[\ie_s^{m_1-1}\overline{\iez}_s^i|v_0] \d s \times \sigma_v,\\
    p_4 &\defeq \int_0^t e^{(j_2+1)ks}\E[\ie_s^{m_1-2}\overline{\iez}_s^{i+1}|v_0] \d s.
\end{align*}
For the special case $m_1=1$, it is easy to find that $\E[\ie_t\overline{\iez}_t^{m_2}|v_0] = 0$.

With the preparations outlined above, the $m$-th conditional central moment of $v_t$ can be calculated as:
\begin{equation*}
    e^{mkt}\E[(v_t-\E[v])^m|v_0] = \sum_{m_1+m_2+m_3=m}\binom{m}{m_1,m_2,m_3}\E[\ie_t^{m_1}\overline{\iez}_t^{m_2}|v_0]\sigma_v^{m_1}(v_0-\E[v])^{m_3}.
\end{equation*}
We note that the conditional central moment $\E[(v_t-\E[v])^m|v_0]$ will be computed as a polynomial in $(v_0-\E[v])$:
\begin{equation*}
    \E[(v_t-\E[v])^m|v_0]
    = c_m(v_0-\E[v])^m + c_{m-1}(v_0-\E[v])^{m-1} + \cdots + c_1(v_0-\E[v]) + c_0,
\end{equation*}
where $c_{m},\dots, c_0$ are coefficients, some of which may be zero and $c_m = e^{-mkt}$. The reason is that the conditional joint moment $\E[\ie_t^{m_1}\overline{\iez}_t^{m_2} | v_0]$ produces a polynomial in $(v_0 - \E[v])$ of order at most $\lfloor m_1 / 2 \rfloor$. We further note that the unconditional central moment of $v_t$ can be computed via:
\begin{equation*}
    \E[(v_t-\E[v])^m] = \E[\E[(v_t-\E[v])^m|v_0]].
\end{equation*}
Meanwhile, due to the assumption that $v_t$ is strictly stationary, we have
\begin{equation*}
    \E[(v_t-\E[v])^m] = \E[(v_0-\E[v])^m].
\end{equation*}
Thus, the $m$-th unconditional central moment of $v_t$ can be computed using the following recursive equation:
\begin{equation}\label{eqn:srjd-moment-recursive}
    (1-e^{-mkt})\E[(v_0-\E[v])^m] = c_{m-1}\E[(v_0-\E[v])^{m-1}] + \cdots + c_1\E[(v_0-\E[v])] + c_0.
\end{equation}
For example, the second central moment is computed as:
\begin{equation*}
    \E[(v_0-\E[v])^2] = \frac{\lambda\mu_v^2}{k} + \frac{\E[v]\sigma_v^2}{2k},
\end{equation*}
and the third central moment:
\begin{equation*}
    \E[(v_0-\E[v])^3] = \frac{2\lambda\mu_v^3}{k} + \frac{\sigma_v^2\lambda\mu_v^2}{k^2} + \frac{\E[v]\sigma_v^4}{2k^2}.
\end{equation*}
Using the recursive Equation~\eqref{eqn:srjd-moment-recursive}, we can compute the fourth and any higher central moments recursively. Given the central moments, the corresponding non-central moments can be easily computed.  

Combining the unconditional moments of $v_0$, computed via Equation~\eqref{eqn:srjd-moment-recursive}, with the polynomial formula in $v_0$ of the conditional moments of the return, computed via Equation~\eqref{eqn:ieii-ieziziz-polynomial}, the unconditional moments of the return of the SVCJ model can be computed. The automation of these derivations has been implemented in the Python package \emph{ajdmom} as well \citep{wu2024ajdmom}.

\end{document}